\documentclass[10pt,aps,prx,twocolumn,superscriptaddress]{revtex4-2}
\usepackage{amsmath, amssymb}
\usepackage{bm}
\usepackage{MnSymbol}
\usepackage{mathtools}
\usepackage[hidelinks]{hyperref}

\newcommand{\bra}[1]{\ensuremath{\left\langle#1\right|}}
\newcommand{\ket}[1]{\ensuremath{\left|#1\right\rangle}}
\newcommand{\braket}[2]{\ensuremath{\left\langle#1|#2\right\rangle}}

\begin{document}

\title{All-optical quantum computing using cubic phase gates}
\author{Niklas Budinger}
\email{nbudinge@t-online.de}
\affiliation{Johannes-Gutenberg University of Mainz, Institute of Physics, Staudingerweg 7, 55128 Mainz, Germany}
\author{Akira Furusawa}
\affiliation{Department of Applied Physics, School of Engineering, The University of Tokyo, 7-3-1 Hongo, Bunkyo-ku, Tokyo 113-8656, Japan}
\affiliation{Optical Quantum Computing Research Team, RIKEN Center for Quantum Computing, 2-1 Hirosawa, Wako, Saitama 351-0198, Japan}

\author{Peter van Loock}
\email{loock@uni-mainz.de}
\affiliation{Johannes-Gutenberg University of Mainz, Institute of Physics, Staudingerweg 7, 55128 Mainz, Germany}

\begin{abstract}

If suitable quantum optical interactions were available,
transforming the field mode operators in a nonlinear fashion,
the all-photonics platform could be one of the strongest contenders
for realizing a quantum computer. 
While single-photon qubits may be processed directly, ``brighter'' logical qubits may be embedded in individual oscillator modes, using so-called bosonic codes, for an in-principle fault-tolerant processing.
In this paper, we show how elements of all-optical,
universal, and fault-tolerant quantum computation can be implemented using only 
beam splitters together with single-mode cubic phase gates in reasonable numbers,
and possibly off-line squeezed-state or single-photon resources.
Our approach is based on a decomposition technique
combining exact gate decompositions and approximate Trotterization. 
This allows for efficient decompositions of certain nonlinear continuous-variable multimode gates into the elementary gates,
where the few cubic gates needed may even be weak
or all identical, thus facilitating potential experiments.
The final gate operations include 
two-mode controlled phase rotation and three-mode Rabi-type
Hamiltonian gates, which are shown to be employable for realizing 
high-fidelity single-photon
two-qubit entangling gates or
creating high-quality Gottesman-Kitaev-Preskill states. 
We expect our method to be of general use
with various applications, including those that rely
on quartic Kerr-type interactions. 

\end{abstract}

\maketitle

\section{Introduction}

The photonics platform offers some clear advantages
to quantum computing in terms of scalability and general error 
robustness, depending on the encoding of the quantum information.
In particular, unlike other, matter-based, 
solid-state or atomic platforms,
photonic qubits can be operated at room temperature
and high clock rates -- as high as GHz or, in principle, even THz.
In addition, recent optical continuous-variable time-domain approaches
are extremely well scalable \cite{FurusawaTimeDomain, AndersenTimeDomain}.

However, there are also two main complications for the universal processing
of photonic qubits or, more generally, quantum optical field modes:
the presence of photon loss and the lack of sufficiently strong
interactions that transform the mode operators in a nonlinear fashion.
In many proposals, the lack of suitable optical nonlinearities 
on the level of the mode operator Hamiltonians is 
circumvented by introducing measurement-induced nonlinearities,
possibly supplemented by an appropriately chosen nonclassical
optical ancilla state \cite{Knill2001, GKP, Bartlett2003}.
A similar, but particularly efficient approach
avoiding large coupling losses is one that,
although still relying upon a nonclassical ancilla state, shifts
part of the nonlinearity into the classical feedforward operations 
\cite{Miyata2016, Konno2021magic, Fcpgexp}.
Nonetheless, on the level of the Hamiltonian of the
field modes interacting with a nonlinear medium, weak cubic mode interactions do occur, and it has been known for long that, in principle, so-called cubic single-mode gates in combination with two-mode beam splitters as well as
single-mode quadratic rotations and linear shifts in phase space lead to
a notion of universal continuous-variable (CV) multimode quantum information
processing \cite{LLoyd1999, BraunsteinRMP2005, Menicucci2006, WeedbrookRMP2012}.

Still one problem remains that these naturally occurring 
cubic nonlinearities, such as three-wave mixing \cite{Langford2011},
generalized ``trisqueezing'' \cite{Braunstein1987, Banaszek1997}, or certain cubic two-mode Hamiltonians \cite{Yanagimoto2022},
are not easy to exploit experimentally
in a loss-tolerant and efficient way, exhibiting a sufficiently strong effective nonlinearity \footnote{Note that in the terminology of nonlinear optics, the cubic and the quadratic Hamiltonians are associated with $\chi^{(3)}$ and $\chi^{(2)}$ interactions, respectively. This is the terminology used in Ref.~\cite{Yanagimoto2022}, while $\chi^{(3)}$ interactions are typically much weaker than $\chi^{(2)}$ interactions. In terms of the mode operators, however, the $\chi^{(3)}$ interaction would be of 4th order like the well-known Kerr interactions and the ``quadratic'' $\chi^{(2)}$ interaction would correspond to a cubic Hamiltonian. The latter is the terminology of the present work. In a way, the ``cubic gates'' as used here thus belong to the class of quadratic nonlinear interactions when the above terminology is applied. Nonetheless, we shall stick to the usual terminology of CV quantum computation \cite{LLoyd1999, BraunsteinRMP2005, Menicucci2006, WeedbrookRMP2012} where the cubic gates represent the lowest-order nonlinear, non-Gaussian operations. Quadratic gates then correspond to the Gaussian gates and quadratic Hamiltonians generate the Gaussian (unitary) transformations.}.
Another problem is that, even when assuming that robust, elementary cubic gates
are experimentally available, the existing schemes require
unrealistically many such cubic gates
of sufficient and variable interaction strength,
which even in a well-scalable time-domain approach would 
become impractical taking into account experimental errors and loss
per physical gate.
Here we address this latter problem and,
assuming that cubic single-mode gates will be
experimentally available \cite{Fcpgexp},
we propose gate decomposition techniques that lead to 
gate sequences of cubic single-mode gates and beam splitters
of the order of ten gates, while demonstrating their use in various
relevant quantum applications.
Moreover, the difficult cubic gates may be chosen
either relatively weak or all identical, thus facilitating potential experimental
implementations.

\begin{figure*}
    \centering
    \includegraphics{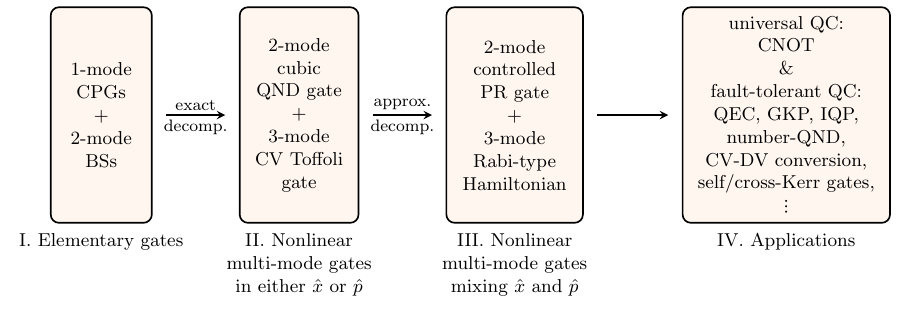}
    \caption{Schematic illustration of the hybrid decomposition method for optical non-Gaussian gates. The acronyms are as follows. CPG, cubic phase gate; BS, beam splitter; QND, quantum nondemolition; PR, phase rotation; QC, quantum computation; QEC, quantum error correction; IQP, instantaneous quantum polynomial \cite{Bremner2016, Douce2017}. Conversions between discrete and continuous variables (CV-DV conversion) can also be achieved with the help of Rabi-type Hamiltonian gates \cite{HastrupConversion2022}. The gates of step II and III are given in equations \eqref{GatesNotMixing} and \eqref{GatesMixing}, respectively.}\label{HDSchema}
\end{figure*}

From a practical point of view, this approach means
that methods to optically realize the simplest,
lowest-order non-Gaussian single-mode CV
operation, which is the cubic phase gate \cite{Miyata2016},
are sufficient to perform all kinds of
advanced optical quantum information processing,
including discrete-variable (DV), entangling gates
on two standard photonic qubits, which are otherwise unavailable
via only linear mode transformations.
Note that it has been shown that naturally occurring
``trisqueezed'' states, which were experimentally already
demonstrated in a nonoptical, superconducting platform \cite{Chang2020}, 
can be converted into cubic phase states,
as typically used as a resource state to implement 
a cubic phase gate, via Gaussian operations \cite{Tri}.

More specifically, we show in this paper how certain elements of all-optical, fault-tolerant, and universal quantum computation can be implemented using primarily Gaussian resources together with cubic phase gates in reasonable numbers. Our approach is based on the efficient decomposition of two non-Gaussian CV multimode gates, namely the two-mode controlled phase rotation gate and the three-mode Rabi-type Hamiltonian gate
\begin{align}\label{GatesMixing}
    e^{i\alpha\hat x_1 \hat n_2}\qquad\qquad\text{and}\qquad\qquad e^{i\beta\hat x_1\hat \sigma_{x, S}},
\end{align}
respectively, where $\hat \sigma_{x, S}$ refers to spin operators as expressed by an optical
Schwinger representation, i.e., one spin represented by two oscillator modes.
As an intermediate set of multimode gates,
we consider so-called two-mode cubic quantum nondemolition (QND) gates
and three-mode CV Toffoli gates
\begin{align}\label{GatesNotMixing}
    e^{i\alpha\hat x_1 \hat x_2^2}\qquad\qquad\text{and}\qquad\qquad e^{i\beta\hat x_1 \hat x_2\hat x_3},
\end{align}
respectively.
These nonlinear multimode gates have the advantage
that they do not mix the two phase-space variables $x$ and $p$,
thus allowing for an exact decomposition into 
single-mode cubic phase gates and beam splitters.
In contrast, the finally obtained nonlinear multimode
gates, as needed for our examples of important
elements in photonic quantum information processing,
do mix $x$ and $p$, and so these require an 
additional approximation step.

Overall, our efficient decomposition method
(see Fig.~\ref{HDSchema}) then relies on a combination of exact decomposition techniques and approximate Trotterization
-- a kind of hybrid decomposition technique, which
we show works remarkably well. 
We analyze the performance of these hybrid decompositions for a single-photon two-qubit controlled-$Z$ gate and for two distinct variants of an optical generation of various manifestations of Gottesman-Kitaev-Preskill (GKP) states
\cite{GKP}.
Generally, we expect our decomposition method to be
of potential use in various other applications, including those 
based on Kerr-type interactions.
One example is a Kerr-interaction-based photon-number QND measurement in a completely
transparent photon detector \cite{Imoto1985, Drummond1994}, which, however,  
can also be realized directly via a cubic, two-mode controlled phase rotation gate, like that for which we derive efficient decompositions, $e^{i\alpha\hat x_1 \hat n_2}$.

As for the non-Gaussian continuous-variable GKP state examples,
our approach is highly compatible with concepts
of measurement-based quantum computing with continuous-variable
cluster states for which the single-mode cubic phase gate
is the canonical non-Gaussian gate \cite{Menicucci2006, Gu2009}.
In this case, the cubic elements may either be introduced
on the level of the measurements,
allowing to measure observables that are no longer
linear combinations of $x$ and $p$, 
i.e., going beyond Gaussian homodyne measurements,
or on the level of the ``off-line''-prepared cluster state
by replacing some of the squeezed-vacuum-state nodes
by cubic phase states -- 
achieving universal continuous-variable ``on-line'' operations
solely by means of Gaussian homodyne measurements.
In the former case, the non-Gaussian measurements
may be based on the detection of photon numbers,
which is generally a common current approach 
to the engineering of non-Gaussian optical states,
i.e., an approach similar to ``Gaussian boson sampling'' \cite{Lund2014, Hamilton2017}
employing Gaussian squeezed-state resources,
linear optics, and photon counting \cite{Sabapathy2019, Tzitrin20}.

Similarly, non-Gaussian states may then be directly
realized within a Gaussian cluster state
through photon number measurements \cite{Eaton2022}.
A potentially useful feature of our approach,
based on the lowest-order non-Gaussian states and gates,
could be that in order to engineer these
simplest nonlinear elements, for instance, via 
non-Gaussian photon measurements, full
photon-number resolution of the optical detectors
up to a sufficiently high number may not be needed \cite{Miyata2016},
similar to the results of Ref. \cite{Fukui2022},
which, however, are still based on an approach close to Gaussian boson sampling.

With our approach, canonical cubic phase gates
realized via a CV cluster state
can be directly used to obtain standard GKP qubit states
as well as GKP ``magic states'' within the cluster state,
independent of photon measurements
beyond the level of, for instance, cubic-phase-state generations.
This would allow to obtain logical non-Clifford gates
for GKP qubits via magic gate teleportation
and nonlinear feedforward \cite{Konno2021magic},
despite the recent result that a single
CV cubic gate acting directly upon a physical
GKP qubit cannot provide a near-unit fidelity non-Clifford gate operation \cite{Hcpg}.
Alternatively, nonlinear gates diagonal in the
number operator, such as those based on a quartic Kerr interaction,
may be employed in order to obtain an efficient and robust
non-Clifford gate for GKP qubits \cite{Royer2022}.
For this type of quartic gate our decomposition 
method into cubic and quadratic gates can also be used.

The paper is structured as follows.
In Sec.~\ref{OQC}, we review the most important elements of the different approaches to optical quantum computing with a particular focus on single-photon qubits and GKP qubits. However, the latter will be introduced in a little more detail in the later section on applications of our methods, Sec.~\ref{Appl}. Our methodology itself will be described in Sec.~\ref{HybridDecompos}.
Sections~\ref{Loss} and \ref{Concl} include a brief discussion of the effects of photon loss on our schemes and a conclusion, respectively.
Two extra appendices present more details on the calculations and the parameter optimizations.

\section{Optical Quantum Computation}\label{OQC}

Photons are robust to decoherence;
however, they get easily lost,
being reflected into the wrong path  
or even absorbed by the environment.
Moreover, for photonic two-qubit gates
based on standard photonic qubits, such as
a CNOT or CZ gate, the necessary nonlinear
interaction $\sim\pi \hat n \otimes \hat n$
is hard to obtain.
In this section, we briefly review notions of universality and
fault tolerance in the context 
of optical encodings and quantum error correction codes, 
including sophisticated
``hardware-efficient'', highly nonclassical ``bosonic codes''.
In the CV setting, for processing
quantum oscillators or ``qumodes'', we briefly discuss
known decomposition techniques as well as optical
approaches to cubic-phase state and gate implementations
as the lowest-order nonlinear resources to introduce a 
non-Gaussian element and complete the universal gate sets.

\subsection{Universal quantum computation}\label{OQCuqc}

Independent of a physical realization,
there are various choices for encoding and processing
quantum information. The most common one is 
that based on qubits and qudits, commonly referred to as DV approach.
Another one is that exploiting continuous quantum variables,
as, for instance, given in a quantized harmonic oscillator,
typically referred to as a ``qumode''. Such latter schemes, processing 
qumodes, are also known as CV quantum information processing
or computing. In either approach, DV or CV, distinct models 
of universal quantum computing have been proposed, most notably
the circuit model \cite{NielsenChuang, LLoyd1999}
based on a reversible sequence of unitary gate operations
and the measurement-based model \cite{Raussendorf2001, Menicucci2006}
based on an irreversible sequence of measurements performed
on a universal, so-called cluster state.

In the context of DV quantum computing with qubits,
the most common universal gate set, approximating, in principle,
any multiqubit operation to arbitrary precision, is
\begin{align}
	\left\lbrace \hat H, \hat S, \hat T, \text{CNOT} \right\rbrace \,,
\label{DVUgateset}\end{align}
where $\hat H$ is the Hadamard gate with 
$\hat H |0\rangle = (|0\rangle + |1\rangle)/\sqrt{2}$,
$\hat H |1\rangle = (|0\rangle - |1\rangle)/\sqrt{2}$,
and the other two single-qubit gates, 
$\hat S = \exp(-i \pi \hat Z/4)$ and
$\hat T = \exp(-i \pi \hat Z/8)$, lead to rotations around the $Z$ axis
of the Bloch representation where $\hat Z$ is a Pauli operator.
The two-qubit CNOT gate applies a bit flip on the second, target qubit only when the first, control qubit is in the state $|1\rangle$.

Note that the set of Eq.~\eqref{DVUgateset}
contains a redundant gate, $\hat S=\hat T^2$.
It is, however, convenient to keep the gate $\hat S$
in the universal set, because it allows to complete
the set of so-called Clifford gates
$\left\lbrace \hat H, \hat S, \text{CNOT} \right\rbrace$, which are known to 
be efficiently simulable by a classical computer \cite{NielsenChuang}.
Towards implementations and fault tolerance, it is useful to reserve the $\hat T$ gate
only for the non-Clifford part of a quantum computation,
which is the crucial part to accomplish universality and 
to circumvent classical simulability for a ``quantum advantage''.
Not to waste non-Clifford gates for Clifford quantum computing
becomes particularly striking in the CV case. 

In this CV case, the most common universal gate set
to process a multiqumode system and obtain a notion of universal
CV quantum computing is
\begin{align}\label{canonCVunivgateset}
	\left\lbrace \hat F, e^{i t \hat x}, e^{i s \hat x^2},
	e^{i r \hat x^3}, \text{CSUM} \right\rbrace \,.
\end{align}
Using the convention $\hbar=1$ throughout this paper, the Fourier gate $\hat F$ is given by $\hat F = \exp\left(i\frac{\pi}{2} \frac{\hat x^2 + \hat p^2}{2}\right) $, allowing to switch between the position and momentum
variables $\hat x$ and $\hat p$, similar to a qubit Hadamard gate.
The other single-qumode gates are $e^{i t \hat x}$, generating momentum shifts,
and the quadratic and cubic phase gates, $e^{i s \hat x^2}$ and $e^{i r \hat x^3}$,
respectively.
The Gaussian quadratic gate involves a single-qumode phase rotation and ``squeezing''.
The two-qumode gate $\text{CSUM} = e^{-i \hat x_1 \hat p_2}$
plays the role of a CV entangling gate,
analogous to the two-qubit $\text{CNOT}$,
transforming the second, target qumode as $\hat x_2 \rightarrow \hat x_2 + \hat x_1$,
the first, control qumode as $\hat p_1 \rightarrow \hat p_1 - \hat p_2$,
while $\hat x_1$ and $\hat p_2$ remain invariant.

Note that both $\hat F$ and $\text{CSUM}$ are not diagonal in the
$x$ variable, whereas all the other gates are.
The latter is useful
to construct a model for CV measurement-based quantum computing
with CV cluster states \cite{Menicucci2006}
and consequently the $\text{CSUM}$ gate is also commonly replaced by
$\text{CZ} = e^{i \hat x_1 \hat x_2}$.
The Fourier gate occurs naturally in CV cluster-state quantum computing
via the elementary teleportations in the cluster state. 

Similar to the discussion above for qubits,
the set of Eq.~\eqref{canonCVunivgateset} also contains a redundant gate, the quadratic phase gate $e^{i s \hat x^2}$, 
which is obtainable from the cubic phase gate
$e^{i r \hat x^3}$ \footnote{Explicitly, this can be seen as follows:  
$e^{it^2\hat x^2}=e^{-i\frac{t^4}{27}}e^{i\frac{t}{3}\hat p}e^{it\hat x^3}e^{-i\frac{t}{3}\hat p}e^{-it\hat x^3}e^{-i\frac{t^2}{3}\hat x}$,
using the momentum-operator Heisenberg transformation 
$e^{-i t \hat x^3}\hat p e^{i t \hat x^3}=\hat p + 3 t \hat x^2$,
as introduced in Sec.~\ref{cubicphasegatesubsection}, and one of the well-known 
Baker-Campbell-Hausdorff (BCH) formulas.},
as pointed out in Refs.~\cite{FurusawaVanLoock, Sefi2011}.
However, it is preferred to keep the quadratic gate in the universal set,
because it allows to complete the ``Clifford set''
for continuous variables,
$\left\lbrace \hat F, e^{i t \hat x},
	e^{i s \hat x^2}, \text{CSUM} \right\rbrace$.
This nonuniversal gate set allows 
to perform any multiqumode Gaussian operation,
corresponding to all linear mode operator transformations
generated by an arbitrary quadratic multimode Hamiltonian.
Similar to the qubit case, there are efficient classical
representations to simulate the Gaussian evolution of Gaussian 
multimode states \cite{Bartlett2002}.

Especially in the optics context, it is better to employ
the non-Gaussian cubic gates as little as possible,
and hence the Gaussian processing should be done
entirely independent of cubic or any higher nonlinear gates.
This minimal use of only the simplest nonlinear gate operations
in optical implementations is one
of the main motivations for the models in the present paper.
Next, we shall briefly discuss the notions of quantum error correction
and fault tolerance, before looking at the most common
ways to combine all these abstract, implementation-independent concepts
in all-optical, universal, and fault-tolerant
quantum computing.

\subsection{Quantum error correction and fault tolerance}

The physical CV errors are continuous and hence the CV states
can be subject to very small, diffusive errors.
In fact, the most common and practically relevant qumode errors
are Gaussian errors such as excitation loss or thermal noise.
In the optics context, excitation loss means photon loss.
Such Gaussian errors generally cannot be suppressed by employing quantum error correction codes that are solely based upon Gaussian states
and operations \cite{Niset2009, Namiki2014}.
	
In the DV approach,	sufficiently many physical qubits can be used to form one logical qubit and for the most common quantum error correction codes, known as stabilizer codes, the Clifford gate set $\left\lbrace \hat H, \hat S, \text{CNOT} \right\rbrace$
is sufficient to construct the logical qubit states as well as to perform the error correction.
In the CV setting, however, fault tolerance and effective quantum error correction
require an additional discretization of the qumode.
A very powerful and prominent code to achieve this is the so-called 
Gottesman-Kitaev-Preskill (GKP) code, encoding a logical qubit in a physical qumode.
This approach is not only ``hardware efficient'', directly
making use of the infinite-dimensional oscillator Hilbert space with no need for adding
extra auxiliary states or modes unless the GKP
qubit code is concatenated with standard multiqubit codes for an enhanced error robustness
\cite{Schmidt2022, Conrad2022, Royer2022, Haenggli2020, Noh2020, Vuillot2019, Fukui2018, Fukui2017}.
It also circumvents the no-go results on Gaussian CV quantum error correction,
enabling one to detect small diffusive errors and correct them
at the expense of a logical error, which then requires an additional higher-level multi-GKP-qubit stabilizer code.
The syndrome measurement of the GKP code is a non-Gaussian operation,
projecting on the GKP code and error spaces, which can be implemented by Gaussian operations together with the non-Gaussian GKP qubit ancilla states.

Thus, bosonic quantum error correction codes
\cite{Grimsmo2020, Michael2016, Bergmann2016cat, Bergmann2016noon, Leghtas2013, Leung1997},
and among them, in particular, the GKP code \cite{GKP, Grimsmo2021}, are an efficient means to protect quantum information embedded into a discretized code space
against CV errors of the physical qumodes.
The GKP code is resource efficient using only a single qumode
and it only requires Gaussian operations for entangling and encoding
qubits, which is of particular practical significance in optics
for the photonic GKP code.
It was shown recently that an extra non-Gaussian element,
beyond that given by a supply of GKP qubits,  
is not even needed for full multiqubit universality
\cite{Yamasaki2020, Baragiola2019}.
Later, as one possible application of our CV gate decompositions
in the context of all-optical implementations,
we will see that the non-Gaussian cubic phase gate $e^{i r \hat x^3}$,
together with some initial Gaussian states, Gaussian homodyne measurements,
Fourier, and beam-splitting operations, allows to generate GKP qubits.
For the nonlinear cubic gate to be experimentally available, 
it may have to be weak.
For it to be robustly implementable, it should not depend
on its continuous, fully tunable operation.
Preferred is a fixed cubic gate with a fixed interaction strength,
for reasons that we discussed in this subsection.
We shall address these issues in our GKP generation scheme.

\subsection{Photonic codes and gates}

The conceptually simplest way to encode a photonic qubit is to only make use of a two-dimensional subspace
of an optical mode's Hilbert space that is spanned by the vacuum
and the single-photon states. More common and convenient than qubit
superposition states of $\left|0\right>$ and $\left|1\right>$
in a single optical mode (single-rail) is to construct a qubit subspace
$\{ \left|1\right>\left|0\right>, \left|0\right>\left|1\right>\}$
on two optical modes (dual-rail). 
Typically, polarization or temporal modes are employed for such dual-rail
photonic qubits.
While all single-qubit gates on photonic dual-rail qubits 
can be realized via linear mode transformations of the two modes,
a two-qubit entangling gate cannot.
Thus, from the universal gate set
$\left\lbrace \hat H, \hat S, \hat T, \text{CNOT} \right\rbrace$,
only $\text{CNOT}$ is hard to obtain directly requiring a quartic, Kerr-type
$\sim\pi \hat n \otimes \hat n$ interaction of sufficient strength $\sim\pi$. 
Applying the corresponding unitary gate $e^{i \pi \hat n \otimes \hat n}$
upon two optical modes with states
$\left|0\right>\left|0\right>$, $\left|1\right>\left|0\right>$,
$\left|0\right>\left|1\right>$, or $\left|1\right>\left|1\right>$
only gives a sign flip for the input state
$\left|1\right>\left|1\right>$ and otherwise acts as the identity
-- a CZ gate. Together with 1-qubit Hadamard gates,
on two-mode qubits simply implementable as a beam splitter,
one can construct a CNOT gate from this. 
The CZ then acts upon modes 2 and 4 of the two dual-rail input qubits
encoded into modes 1, 2 and 3, 4, respectively. 

Thus, efficient photonic quantum computation based on single-photon qubits, each encoded into two modes,
if directly implemented in a unitary circuit model, 
would depend on the availability of a robust, loss-tolerant,
sufficiently strong, nonlinear two-mode gate.
An alternative is a single-mode, nonlinear or Kerr-type gate 
$\sim\pi \hat n^2$ of similar strength $\sim\pi$ in combination
with beam splitters \cite{Knill2001, Ewert2019}, which can either provide
a near-deterministic, heralded nonlinear single-mode gate or
correspond to a, in principle deterministic, quartic CV 
single-mode interaction gate.
We will see that our CV gate decompositions allow to obtain
photonic two-qubit entangling gates from single-mode cubic phase gates.

Probabilistic nonlinear gates can be combined with gate teleportation techniques.
In fact, to circumvent the ``on-line'' implementation of a photon-photon CNOT gate,
measurement-based schemes have been proposed
making use of multiphoton ancilla states \cite{Knill2001},
for instance, in the form of cluster states
\cite{Raussendorf2001, Nielsen2004, Browne2005}.
The entangling gates to build a sufficiently large cluster state ``off-line''
may then be probabilistic.
Similar approaches can be used in order to incorporate 
quantum error correction codes and a notion of loss and even fault tolerance
into the schemes \cite{Varnava2006}.
Nonetheless, for a large-scale quantum computer, besides the experimental
complication of multiplexing and at least short-term storage of quantum information,
a large resource overhead is expected.
In a DV time-domain approach, high experimental source clock rates, for example using quantum dots, are a promising element 
to create this overhead in a practical fashion.

A more direct approach would be based on processing continuous variables, i.e., the mode's
continuous degrees of freedom. In quantum optics this is typically done by relating the in- and out-of-phase amplitudes of the electromagnetic field to the quadrature operators $\hat x$ and $\hat p$. Then
the Gaussian gates, 
$	\left\lbrace \hat F, e^{i t \hat x}, e^{i s \hat x^2},
	\text{CSUM} \right\rbrace$,
can be efficiently realized experimentally.
The quadratic phase gate can be replaced by an optical, single-mode squeezing
operation
$e^{is \left( \hat x \hat p +\hat p\hat x\right)}$
and CSUM may be substituted by an optical beam splitter
$e^{i\theta \left( \hat p_1\hat x_2 - \hat x_1 \hat p_2\right)}$
as it is decomposable into beam splitters
and single-mode squeezers \cite{Braunstein2005}.
In contrast, the non-Gaussian, single-mode cubic phase gate to achieve
CV universality $e^{i r \hat x^3}$ is more difficult to obtain,
requiring a nonlinear optical mode transformation.
We shall address this in Sec.~\ref{cubicphasegatesubsection}.
Note that the Gaussian entangling gates, CSUM or a beam splitter,
are, unlike the CNOT gate
for single-photon qubits, relatively easy to implement. Other optical CV encodings exist as well, most notably utilizing the continuous time-frequency degrees of freedom. Here, non-Gaussian operations as well as GKP states have been experimentally demonstrated \cite{RefereeP1, RefereeP2}; however, performing interactions between modes remains challenging, exhibiting a closer resemblance to the DV than to the CV approaches described above.

Conceptually different from a direct processing of logical,
CV quantum information is to encode logical DV states or especially qubits
into physical, optical CV systems.
This leads to new possibilities of photonic quantum information
processing, in particular, in the context of quantum error correction, 
but also to new types of complications.
Below, when discussing various applications of our approach, we will consider the GKP code, which as an instance of
a shift-invariant bosonic code is an example of such a photonic code.
Its optical manipulation is based on a translation
of the CV quantum optical gate operations
into logical gates acting on the GKP code space.
This results, in principle, in a higher level 
of scalability, especially when the modes to be processed are primarily 
defined in the time domain \cite{FurusawaTimeDomain, AndersenTimeDomain}.
The GKP states are still hard to obtain on demand, but nonetheless
allow for a loss- and fault-tolerant processing that
goes beyond simple quantum error detection and allows for photonic quantum error correction.
Particularly attractive is that the use of
efficient Gaussian two-mode gates for entangling 
two qumodes from the CV setting directly translates
to GKP qubits and their logical CNOT gates.
However, single-qubit universality for GKP qubits
is harder than that for single-photon qubits,
requiring nonlinear mode operator transformations. 
Next let us take a look at the most common methods 
for gate decomposition.

\subsection{Gate decomposition techniques}

In the year 1999, Lloyd and Braunstein demonstrated that every multiqumode operation could, in principle, be approximated to arbitrary precision using only gates from the universal gate set 
of Eq.~\eqref{canonCVunivgateset} \cite{LLoyd1999}. 
However, for the experimental feasibility of a given operation, the number of elementary gates required for its approximation is no less important. Consequently, various CV gate decomposition schemes have been developed enabling certain groups of qumode operations to be implemented more efficiently. We shall briefly discuss the Trotter-Suzuki decomposition \cite{SUZ1, SUZ2}, the commutator-based approach proposed by Lloyd and Braunstein \cite{LLoyd1999} and optimized in Ref.~\cite{Sefi2011}, as well as the exact gate decomposition scheme \footnote{The possibility and usefulness of an exact decomposition for some given nonlinear gates were first pointed out in Refs.~\cite{Sefi2011, Sefi2013}. A more systematic treatment, exploring more generally which gate classes may be exactly decomposed, was presented later in Ref.~\cite{Aed}.} as developed in Ref.~\cite{Aed}. An efficient and exact decomposition of Gaussian operations can be found in Ref.~\cite{Ukai2010} adapted to CV cluster computation.

The Trotter-Suzuki decomposition can be used to obtain the operation $e^{it\left(\hat A+\hat B\right)}$ from the two gates $e^{it\hat A}$ and $e^{it\hat B}$. It relies upon the Lie-Trotter product formula,
\begin{align}
	\left(e^{i\frac{t}{n}\hat A}e^{i\frac{t}{n}\hat B}\right)^{n}\stackrel{n\rightarrow\infty}{\longrightarrow}e^{it\left(\hat A+\hat B\right)}.
\end{align}
By introducing and adapting individual gate strengths $t_i$ of the $n$ repetitions, the order of convergence can be chosen arbitrarily high as shown in Refs.~\cite{SUZ1, SUZ2}.
Similarly, in the commutator-based decomposition schemes \cite{LLoyd1999, Sefi2011}, the operator $e^{t\left[\hat A, \hat B\right]}$ is approximated using the relation
\begin{align}
	\left(e^{-i\sqrt{t/n}\hat A}e^{-i\sqrt{t/n}\hat B}e^{i\sqrt{t/n}\hat A}e^{i\sqrt{t/n}\hat B}\right)^{n}\stackrel{n\rightarrow\infty}{\longrightarrow}e^{t\left[\hat A, \hat B\right]},
\end{align}
given the operators $e^{it\hat A}$ and $e^{it\hat B}$. Repeating this procedure then also enables the creation of operations with nested commutators. In combination with the Trotter-Suzuki decomposition and the universal gate set, this then allows for the implementation of any polynomial of the bosonic mode operators $\hat x_i$ and $\hat p_i$ and hence any multi-qumode operation \cite{LLoyd1999}. Again, by adapting the gate strengths $t_i$ of the different repetitions, the order of convergence can be increased which significantly enhances the efficiency of the decomposition when targeting large gate strength $t$ along with high accuracies. Nevertheless, the number of approximate steps needed to arrive at the desired operation lets the amount of required elementary gates rise rapidly. Therefore, whenever possible the exact decomposition of gates is preferable.

The exact gate decomposition scheme by Kalajdzievski and Arrazola \cite{Aed} enables the decomposition of operators of the general form
\begin{align}
	\exp\Big(it\left({\textstyle\prod_{j=1}^{N-1}}\hat x_j\right)\hat x_N^n\Big),\label{ExactlyDecomposable}
\end{align}
where $N$ as well as $n\cdot N$ must be divisible by either two or three. As this is based on the relation
\cite{Aed, Sefi2013, Sefi2011}
\begin{align}
	e^{i\alpha\hat x_j^m\hat p_k}e^{it\hat x_k^n}e^{-i\alpha\hat x_j^m\hat p_k}=e^{it\left(\hat x_k+\alpha\hat x_j^m\right)^n}
\end{align}
and the eventual cancellation of unwanted polynomial terms of the right-hand side, the optical position and momentum operators $\hat x_i$ and $\hat p_i$ must not be mixed. Hence, the subgroup of exactly decomposable gates is rather small. Nevertheless, if a given quantum operator is in the form of Eq.~\eqref{ExactlyDecomposable}, its exact decomposition is generally superior to approximate approaches in gate count and accuracy.

\subsection{Optical cubic phase states and gates}\label{cubicphasegatesubsection}

The hardest gate of the universal gate set of 
Eq.~\eqref{canonCVunivgateset} to realize quantum optically is the single-mode cubic phase gate \cite{GKP}, $e^{i r \hat x^3}$, which is the only non-Gaussian gate of the set. 
It acts trivially on the position operator, $e^{-i r \hat x^3} \hat x e^{i r \hat x^3} = \hat x$, while the momentum operator is transformed by it in a nonlinear fashion \cite{FurusawaVanLoock},
\begin{align}\label{CPGHeisenberg}
	e^{-i r \hat x^3} \hat p \, e^{i r \hat x^3} = \hat  p + 3 r \hat x^2,
\end{align}
shifting the momentum by the squared position
\footnote{Here we use the convention $\hbar = 1$ throughout. Generally, one has
$e^{-i r \hat x^3} \hat p \, e^{i r \hat x^3} = \hat p + [\hat p, ir \hat x^3] = \hat p + 3 \hbar r \hat x^2$
using the BCH formula $e^{-B} A e^{B}= A + [A,B]$ (where the additional nested commutators of the sum have vanished) with $[\hat x^3, \hat p] = 3 i \hbar \hat x^2$.}.
This gate was originally introduced as one option to add a non-Clifford element and complete the logical universal gate set for GKP qubits \cite{GKP}. We will come back to this later.
While the gate is also the canonical choice to achieve CV universality, it is particularly well suited to incorporate a non-Gaussian element into the concept of CV cluster computation \cite{Menicucci2006, Gu2009}. Earlier it was considered in another variant of measurement-based quantum computation, namely a version of CV optical gate teleportation \cite{Bartlett2003}.

The original idea to optically obtain a cubic phase gate was based on photon measurements on parts of Gaussian states \cite{GKP, Gu2009, Sabapathy2019}, which conditionally prepares a cubic phase state $\int \hspace{-0.7mm}\text{d}x \,e^{i r x^3}|x\rangle$ that can be used as a resource for cubic-phase-gate teleportation. Alternatively, small-number Fock superposition states, being approximations of cubic phase states, may be directly used as single-mode ancilla states for weak cubic-phase-gate teleportation \cite{Yukawa2013}. Later, the concept of nonlinear squeezing was introduced, which is related to the non-Gaussian and nonclassical properties of a nonlinear, cubic phase state \cite{Konno2021}. The higher the nonlinear squeezing, corresponding to a higher average number of photons, the stronger the cubic phase gate becomes that can be obtained with the help of the approximated cubic phase state. Eventually, it was shown that nonlinear feedforward operations based on the results obtained from homodyne detectors allow to effectively measure nonlinear quadrature combinations. Combined with the nonclassical photon ancilla states, this offers an efficient way to optically realize a single-mode cubic phase gate \cite{Miyata2016}. The technique of nonlinear feedforward can also be employed to achieve magic gate teleportation using magic states for GKP qubits \cite{Konno2021magic}. 

Most recently, the nonlinear feedforward operation for cubic phase gate teleportation was experimentally demonstrated \cite{Fcpgexp}. The degree of nonlinear squeezing of the non-Gaussian ancilla state experimentally achieved is related to the cubic phase gate strength parameter $r$ in Eq.~\eqref{CPGHeisenberg} as $r \approx 0.17$. An ancilla state with a higher photon number would exhibit larger nonlinear squeezing, and this can be used to make the resulting cubic phase gate stronger. We will show that our decomposition method, when applied and optimized for GKP state generation of fairly high fidelity above 90\% using single-mode cubic phase gates and beam splitters, yields 
parameter values all below $r = 0.17$. Thus, our decomposition-based scheme is fully compatible with the recent experimental nonlinear-feedforward demonstration. In order to achieve better fidelities, we would apply a larger sequence of gates where each gate is typically even weaker. The remaining complication in a practical application of these schemes would be loss and noise as well as relatively low gate fidelities, so that a smaller sequence of imperfect gates is preferable. We leave a complete analysis of the experimental scheme of Ref.~\cite{Fcpgexp}, including loss and imperfections, applied to our gate decompositions for, especially, GKP state generation to future work. The important conclusion here is that the methods of Refs.~\cite{Yukawa2013, Miyata2016, Konno2021, Fcpgexp, Konno2021magic} can be very well combined with our present approach.

\section{Hybrid Decomposition Scheme for optical non-Gaussian Gates}\label{HybridDecompos}

The basis of all subsequent considerations is the universal gate set given by
\begin{align}
	\left\lbrace e^{i\pi \left( \hat x^2 + \hat p^2\right) }, e^{it \hat x}, e^{is \left( \hat x \hat p +\hat p\hat x\right)}, e^{i\theta \left( \hat p_1\hat x_2 - \hat x_1 \hat p_2\right) }, e^{ir \hat x^3} \right\rbrace 
\end{align}
with $t, s, \theta, r \in \mathbb{R}$ and the quadrature operators $\hat x_k=\frac{1}{\sqrt{2}}\left( \hat a_k + \hat a_k^\dagger\right)$ and $\hat p_k=\frac{1}{\sqrt{2}i}\left( \hat a_k - \hat a_k^\dagger\right)$. The Gaussian operations, namely phase rotation, displacement, squeezing, and beam splitting are all readily available in experimental quantum optics, as discussed in Sec.~\ref{OQC}. The cubic phase gate, on the other hand, is experimentally more challenging to implement, but also for this very recent demonstrations exist, as described in Sec.~\ref{cubicphasegatesubsection}.

Using only gates from this set we want to approximate two multimode gates, the controlled phase rotation gate $e^{i\alpha\hat x_1\hat n_2  }$ and the Rabi-type-Hamiltonian gate $e^{i\beta\hat x_1 \hat \sigma_{x, S}}$, where the latter refers to spin operators as expressed by an optical Schwinger representation,
i.e., one spin represented by two oscillator modes. Both types of gates can be used to achieve universal quantum computing in an optical setting, as we will see later.
As the given gate set is universal, it is possible to approximate any multiqumode quantum gate using only a finite number of gates from the set.  However, the experimental feasibility of any gate approximation depends heavily on the amount of concatenated gates. Hence the focus of this paper lies on obtaining good approximations while also minimizing the number of basic operations needed.\\
To achieve this, we use a hybrid decomposition approach, which is based on exact decomposition schemes followed by one step of the approximation technique known as Trotterization,
as illustrated in Fig.~\ref{HDSchema}. This way, we are able to decompose nonlinear multimode gates, which mix $\hat x$ and $\hat p$ while still utilizing the low gate counts and unit fidelities of exact decompositions.
While the general efficiency of a gate decomposition is difficult to quantify as it depends on the specific application as well as the required accuracy, we presume this combination of exact decompositions with only one approximate step to be generally more efficient than commutator-based approximations.
Let us now discuss these individual techniques.

\subsection{Exact decomposition of cubic multimode gates}

The first step of the presented approximation scheme is the exact decomposition of two cubic multimode gates, namely the cubic QND gate $e^{i\alpha \hat x_1 \hat x_2^2}$ and the CV Toffoli gate $e^{i\beta \hat x_1 \hat x_2 \hat x_3}$. These gates will be the basic elements of the following Trotterization.
Using a lemma to the Baker-Campbell-Hausdorff formula,
\begin{align}
	e^{\hat A} \hat B e^{-\hat A} = \sum_{n=0}^\infty \frac{1}{n!} \bigl[\underbrace{\hat A, \left[\hat A, ...\left[\hat A\right.\right.}_n\left.\left., \hat B\right]...\right]\bigr]\,,
\end{align}
and the beam splitter $\hat B_{k l}(s)=e^{i s\left( \hat p_k \hat x_l - \hat x_k\hat p_l \right)}$, we easily obtain
\begin{align}
	\hat B_{12}(s)\hat x_1 \hat B_{12}(-s) = \cos(s)\hat x_1 +\sin(s)\hat x_2\,,
\end{align}
and thus
\begin{align}\begin{split}
	&\hat B_{12}(s)e^{ir\hat x_1^3}\hat B_{12}(-2s)e^{ir\hat x_1^3}\hat B_{12}(s)\\
	=& e^{ir(\cos(s)\hat x_1 +\sin(s)\hat x_2)^3}
	\thinspace e^{ir(\cos(s)\hat x_1 -\sin(s)\hat x_2)^3}\\
	=& e^{2ir\cos^3(s)\hat x_1^3}e^{6ir\cos(s)\sin^2(s)\hat x_1\hat x_2^2}.
\end{split}\end{align}
This leads us to the exact decomposition of a cubic QND gate,
\begin{align}
	e^{i\alpha\hat x_1\hat x_2^2}=\hat B_{12}(s)e^{ir\hat x_1^3}\hat B_{12}(-2s)e^{ir\hat x_1^3}\hat B_{12}(s)e^{-i\beta\hat x_1^3},\label{decomp1}
\end{align}
with $\alpha = 6r\cos(s)\sin^2(s)$ and $\beta = 2r\cos^3(s)$ using a total of three single-mode cubic phase gates and three beam splitters.
Continuing with the result of Eq.~\eqref{decomp1} we directly obtain
\begin{align}\begin{split}
	&\hat B_{23}(u)e^{it\hat x_1\hat x_2^2}\hat B_{23}(-2u)e^{-it\hat x_1\hat x_2^2}\hat B_{23}(u)\\
	=& e^{it\hat x_1(\cos(u)\hat x_2 +\sin(u)\hat x_3)^2}
	\thinspace e^{-it\hat x_1(\cos(u)\hat x_2 -\sin(u)\hat x_3)^2}\\
	=&e^{4it\sin(u)\cos(u)\hat x_1 \hat x_2\hat x_3}=e^{2it\sin(2u)\hat x_1 \hat x_2\hat x_3}.
\end{split}\end{align}
Note that the last term of Eq.~\eqref{decomp1} can be omitted in the above calculation. Hence we get a CV Toffoli gate at the expense of four single-mode cubic phase gates and nine beam splitters. In comparison to existing exact decomposition schemes \cite{Aed} this not only reduces the number of especially single-mode cubic phase gates, but it also works using only simple beam splitters instead of the experimentally more challenging quadratic QND gates that involve additional squeezing operations.
Table~\ref{exactdecomptable} lists the number of elementary gates needed for the exact decompositions of \cite{Aed}.\begin{table}
	\centering
        \caption{Comparison of the number of elementary gates needed for the exact decomposition of the cubic QND gate $e^{i\alpha\hat x_1 \hat x_2^2}$ and the CV Toffoli gate $e^{i\beta\hat x_1 \hat x_2\hat x_3}$ with the corresponding gate counts from Kalajdzievski \textit{et al}. \cite{Aed}. The quadratic QND gates required in \cite{Aed} are either CZ or CSUM gates.}\label{exactdecomptable}
        \begin{ruledtabular}
	\begin{tabular}{lcccc}
        & \multicolumn{2}{c}{Present paper} & \multicolumn{2}{c}{Kalajdzievski \textit{et al}.} \\[1mm]\cline{2-3}\cline{4-5}
		& Cubic phase & Beam & Cubic phase & QND \\[-0.7mm]
        Target gate & gates & splitters & gates & gates \\[1mm]\hline
		Cubic QND & 3 & 3 & 5 & 4 \\
		CV Toffoli & 4 & 9 & 7 & 10
	\end{tabular}
        \end{ruledtabular}
\end{table}
Note that the name CV Toffoli gate is usually assigned to the gate $e^{i\beta\hat x_1 \hat x_2\hat p_3}$ instead of the gate $e^{i\beta\hat x_1 \hat x_2\hat x_3}$ used in this paper. Analogous to a DV three-qubit Toffoli gate, which applies a bit flip to the third qubit only when the two first qubits are both in the logical state $\left|1\right>$, this commonly defined CV Toffoli gate shifts the position of mode 3, $\hat x_3$, by an amount proportional to the product of the positions of modes 1 and 2, $\hat x_1 \hat x_2$. However, both definitions only differ by a Fourier gate, which is easily implemented optically.

\subsection{Efficient Trotter-Suzuki decomposition of the controlled phase rotation gate}

\begin{figure*}
    \centering
    \includegraphics{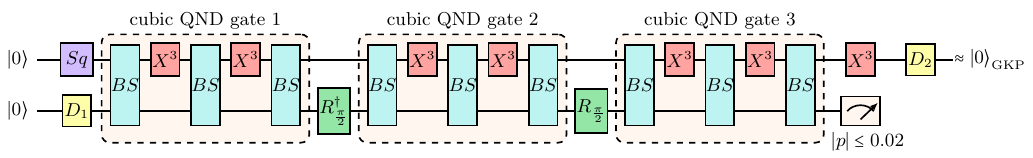}
    \caption{Exemplary implementation of the presented hybrid decomposition scheme. Three exactly decomposed cubic QND gates are concatenated and their gate strengths are optimized based on the Trotter-Suzuki decomposition. Using an off-line squeezing of 12.2 dB (violet), two displacements (yellow), two Fourier gates (green), nine beam splitters (blue), seven cubic phase gates (red), and a conditional homodyne measurement with a success probability of 0.2\% gives a $\left|0\right>_\text{GKP}$ state with a fidelity of $\geq$90.5\%. Explicit gate strengths can be found in Appendix~\ref{AppOPS}.}\label{Schema}
\end{figure*}

Concatenating several cubic QND gates, while Fourier transforming the second mode of every second gate, we obtain
\begin{align}
	S_{\bm{\lambda}}(t) = \prod_{j = 1}^{m}\exp\left({it\lambda_j\hat x_1\frac{\hat x_2^2}{2}}\right)\thinspace\exp\left({it\mu_j\hat x_1\frac{\hat p_2^2}{2}}\right),
\end{align} with the indexation of the product going from right to left and $\bm{\lambda}=\left( \lambda_m, \mu_m, ..., \lambda_1, \mu_1\right) ^T$.
The first-order Trotter-Suzuki decomposition is then given by the Lie-Trotter product formula
\begin{align}
	\exp\left({i\alpha\hat x_1\frac{\hat x_2^2 + \hat p_2^2}{2}}\right)=\left[S_{(1, 1)^T}\left(\frac{\alpha}{b}\right)\right]^b + \mathcal{O}\left(\frac{1}{r}\right),
\end{align}
with $b\in\mathbb{N}$. Higher orders of convergence can be achieved by using specific sets of parameters $\bm{\lambda}$ as provided by Suzuki in Refs.~\cite{SUZ1, SUZ2}.
However, for small $b$ the order of convergence should not surpass its role of guidance: As demonstrated in Ref.~\cite{CCts}, finding the right set of parameters for a specific problem instead of using the common Trotter-Suzuki decompositions can significantly enhance the approximation.
But before we can attempt such a parameter optimization, we need to determine the impact of different parameter sets $\bm{\lambda}$ on the different applications. In order to do this analytically, a change in representation will prove to be useful:\\
Let us regard the impact of the operator on the different quadratures of the two modes. Note that, up to a global phase, this defines an arbitrary operator unambiguously (see Appendix~\ref{AppCalculations}).
Starting with the second mode and using $e^{i\frac{t}{2}\hat x^2}\hat p\,e^{-i\frac{t}{2}\hat x^2}=\hat p-t\hat x$ and $e^{i\frac{t}{2}\hat p^2}\hat xe^{-i\frac{t}{2}\hat p^2}=\hat x+t\hat p$, it is easily seen that
\begin{align}\begin{split}
	S_{\bm{\lambda}}^b(t)\hat x_2S_{\bm{\lambda}}^b(-t)&=P_{xx}[t\hat x_1]\hat x_2 + P_{xp}[t\hat x_1]\hat p_2,\\
	S_{\bm{\lambda}}^b(t)\hat p_2S_{\bm{\lambda}}^b(-t)&=P_{px}[t\hat x_1]\hat x_2 + P_{pp}[t\hat x_1]\hat p_2.
\end{split}\end{align}
The thereby defined polynomials $P_{xx}$, $P_{xp}$, $P_{px}$, and $P_{pp}$ are all of the order $2m b\equiv L$ and can easily be calculated for a given $\bm{\lambda}$. The corresponding recursive formulas are presented in Appendix~\ref{AppCalculations}.

While the operator's impact on the first mode's quadratures is slightly more complex, it is also solely dependent on these four polynomials. Consequently, $P_{xx}$, $P_{xp}$, $P_{px}$, and $P_{pp}$ define the approximated controlled phase rotation gate up to a global phase and provide an equal yet far more intuitive representation of $S_{\bm{\lambda}}^b$ than $\bm{\lambda}$ and $b$: 
the operator $S_{\bm{\lambda}}^b$ approximates the controlled phase rotation gate $e^{it\hat x_1\hat n_2}$ with
\begin{align}\begin{split}
	e^{it\hat x_1\hat n_2}\hat x_2e^{-it\hat x_1\hat n_2}&=\cos(t\hat x_1)\hat x_2 + \sin(t\hat x_1)\hat p_2,\\
	e^{it\hat x_1\hat n_2}\hat p_2e^{-it\hat x_1\hat n_2}&=-\sin(t\hat x_1)\hat x_2 + \cos(t\hat x_1)\hat p_2
\end{split}\end{align}
if and only if the four polynomials approximate the four functions cosine, sine, -sine, and cosine, respectively. The best approximation for a given $L$ is thus obtained by simply choosing the truncated Taylor expansions of sine and cosine. Following directly from the recursive formulas is also the common as well as useful relation
\begin{align}\label{sincosinerelation}
	P_{xx}[t]P_{pp}[t] - P_{xp}[t]P_{px}[t] &= 1,\quad \forall t\in\mathbb{R}.
\end{align}

\subsection{Decomposition of the Rabi-type Hamiltonians}\label{Rabidecompos}

The Trotter-Suzuki decomposition of the Rabi-type Hamiltonian gate $e^{i\alpha\hat x_1\hat \sigma_{x, S}}$ works rather similar.
When replacing the cubic QND gates with CV Toffoli gates we obtain the operator
\begin{align}
	T_{\bm{\lambda}}(t) = \prod_{j = 1}^{m}\exp\left({it\lambda_j\hat x_1\hat x_2\hat x_3}\right)\thinspace\exp\left({it\mu_j\hat x_1\hat p_2\hat p_3}\right),
\end{align} 
with the indexation of the product going from right to left and $\bm{\lambda}=\left( \lambda_m, \mu_m, ..., \lambda_1, \mu_1\right) ^T$ as before.
Using modes 2 and 3 as a dual-rail qubit
\begin{align}
    \beta\left|0\right>_\text{DR}+\gamma\left|1\right>_\text{DR}&=\beta\left|1\right>_2\left|0\right>_3+\gamma\left|0\right>_2\left|1\right>_3\,,
\end{align}
together with the Schwinger representation of the Pauli operator
\begin{align}
    \hat\sigma_{x, S}=\hat a_2^\dagger \hat a_3+\hat a_3^\dagger \hat a_2
\end{align}
the Rabi-type Hamiltonian gate can be written as
\begin{align}
	e^{i\alpha\hat x_1\hat \sigma_{x, S}}&=e^{i\alpha\hat x_1(\hat a_2^\dagger \hat a_3+\hat a_3^\dagger \hat a_2)}=e^{i\alpha\hat x_1(\hat x_2\hat x_3+\hat p_2\hat p_3)},
\end{align}
and from the Lie-Trotter product formula it follows that
\begin{align}
	\left[T_{(1, 1)^T}\left(\frac{\alpha}{b}\right)\right]^b&\stackrel{b\rightarrow\infty}{\longrightarrow}e^{i\alpha\hat x_1(\hat x_2\hat x_3+\hat p_2\hat p_3)}=e^{i\alpha\hat x_1\hat \sigma_{x, S}}.
\end{align}
Furthermore, with similar impact on the different quadratures,
\begin{align}\begin{split}
		T_{\bm{\lambda}}^b(t)\hat x_2T_{\bm{\lambda}}^b(-t)&=P_{xx}[t\hat x_1]\hat x_2 + P_{xp}[t\hat x_1]\hat p_3,\\
		T_{\bm{\lambda}}^b(t)\hat p_2T_{\bm{\lambda}}^b(-t)&=P_{px}[t\hat x_1]\hat x_3 + P_{pp}[t\hat x_1]\hat p_2,\\
		T_{\bm{\lambda}}^b(t)\hat x_3T_{\bm{\lambda}}^b(-t)&=P_{xx}[t\hat x_1]\hat x_3 + P_{xp}[t\hat x_1]\hat p_2,\\
		T_{\bm{\lambda}}^b(t)\hat p_3T_{\bm{\lambda}}^b(-t)&=P_{px}[t\hat x_1]\hat x_2 + P_{pp}[t\hat x_1]\hat p_3\,,
\end{split}\end{align}
the same polynomials as before, $P_{xx}$, $P_{xp}$, $P_{px}$, and $P_{pp}$ can be used to define the approximated Rabi-type Hamiltonian gate up to a global phase
(see Appendix~\ref{AppCalculations}).\\
Note that with the Schwinger representation of the remaining Rabi-type Hamiltonian gates, 
\begin{align}
	e^{i\alpha\hat x_1\hat \sigma_{y, S}}&=e^{i\alpha\hat x_1(i\hat a_3^\dagger \hat a_2-i\hat a_2^\dagger \hat a_3)}=e^{i\alpha\hat x_1(\hat x_2\hat p_3-\hat p_2\hat x_3)},\\
	e^{i\alpha\hat x_1\hat \sigma_{z, S}}&=e^{i\alpha\hat x_1(\hat a_2^\dagger \hat a_2-\hat a_3^\dagger \hat a_3)}=e^{i\alpha\hat x_1\frac{\hat x_2^2+\hat p_2^2}{2}-i\alpha\hat x_1\frac{\hat x_3^2+\hat p_3^2}{2}},
\end{align}
it is easily seen that the operators $S_{\bm{\lambda}}(t)$ and $T_{\bm{\lambda}}(t)$ are sufficient to approximate the full set of Rabi-type Hamiltonian gates. More precisely, we have
\begin{align}
	\hat F_3\left[T_{(1, 1)^T}\left(\frac{\alpha}{b}\right)\right]^b\hat F_3^\dagger&\stackrel{b\rightarrow\infty}{\longrightarrow}e^{i\alpha\hat x_1(\hat x_2\hat p_3-\hat p_2\hat x_3)},\\
	\left[S_{(1, 1)^T}^{(1, 2)}\left(\frac{\alpha}{b}\right) S_{(1, 1)^T}^{(1, 3)}\left(-\frac{\alpha}{b}\right)\right]^b&\stackrel{b\rightarrow\infty}{\longrightarrow}e^{i\alpha\hat x_1\frac{\hat x_2^2+\hat p_2^2-\hat x_3^2-\hat p_3^2}{2}},
\end{align}
with the Fourier transform $\hat F_3=\exp(i\frac{\pi}{2}\frac{\hat x_3^2 + \hat p_3^2}{2})$ and the superscripts of $S_{\bm{\lambda}}$ here and in the following denoting on which modes the gates act upon. As an overall result, we have effectively obtained an efficient decomposition of general Rabi-type Hamiltonian gates based on CV Toffoli gates and their exact decompositions into a set of elementary CV operations that solely contains single-mode cubic phase gates and beam splitters.
Though an approximation, in principle, this allows to deterministically simulate a general Rabi-type Hamiltonian interaction by optical means \footnote{This is unlike previous treatments that rely upon non-optical platforms with more natural Rabi-type interactions \cite{Park2017} or similarly on dispersive atom-light interactions in cavityQED \cite{Loock2008}, or, alternatively, in a certain instance of an optical two-mode Rabi-type interaction including an optical ``single-mode qubit'', on probabilistic conditional operations \cite{Park2020}. Our scheme, of course, would still fundamentally depend on the availability of optical single-mode cubic phase gates. In some of the other treatments, the initial setting is somewhat converse compared with ours: Rabi-type qumode-qubit interactions including a physical two-level qubit system are assumed to be available and then concatenated to obtain, for instance, a qumode (e.g., optical) single-mode cubic (or even higher) phase gate \cite{Park2018}.}.

\subsection{Optimizing the Trotter-Suzuki decomposition for different applications}

One advantage of focusing on the four polynomials, as introduced above, is the simplicity with which they allow us to check for the order of convergence of the approximations. As the Taylor expansions of sine and cosine are well known, a simple comparison of all terms of corresponding order is sufficient. Consequently, the order of convergence $n$ of a parameter set $\bm{\lambda}$ is coupled to a system of equations
\begin{align}
	\ \ \qquad\smashoperator{\sum_{0\leq k_1\leq l_1<k_2\leq l_2<...\leq m}}\ \overbrace{\mu_{k_1}\lambda_{l_1}\mu_{k_2}\lambda_{l_2}\cdots}^n=\smashoperator{\sum_{0\leq l_1<k_2\leq l_2<k_3\leq...\leq m}}\ \overbrace{\lambda_{l_1}\mu_{k_2}\lambda_{l_2}\mu_{k_3}\cdots}^n=\frac{1}{n!},
	\label{conditionsOOC}
\end{align}
where the left-hand side gives the two nonzero coefficients of $n$th order of the four polynomials and the right-hand side the corresponding coefficient of sine and cosine.
While the common Trotter-Suzuki decomposition builds upon the symmetric operator $S_{\left(\frac{1}{2}, 1, \frac{1}{2},0\right)}$ resulting in only even orders of convergence, this allows us to also test {\it odd orders}.
For example, the third order, after choosing its one degree of freedom appropriately, was found to be superior to comparable even orders in many of the applications tested in the scope of this paper. On the other hand, leaving the restrictions of Eq.~\eqref{conditionsOOC} behind and conducting a free parameter search significantly improved results further. Therefore, unless stated otherwise, all presented approximations are optimized using the Basin-hopping algorithm as implemented by SciPy with starting points fulfilling Eq.~\eqref{conditionsOOC}, maximizing the fidelity to the respective target state. The explicit sets of parameters $\bm{\lambda}$ can be found in Appendix~\ref{AppOPS}.

The two separate steps of the hybrid decomposition scheme are illustrated in 
Fig.~\ref{Schema}, where three exactly decomposed cubic QND gates are concatenated with optimized gate strengths in order to approximate a GKP state. Such GKP state generations are one of the possible applications of our method, which we shall treat in more detail next.

\section{Applications}\label{Appl}

\subsection{Approximating the qubit CZ gate}

As a first application of our method we are going to look at a controlled-$Z$ gate for photonic qubits, realizable via the two-mode gate $e^{i\pi\hat n_1\hat n_2}$.
While single-qubit operations on photonic qubits are easily implemented experimentally using the gates from the universal gate set, as discussed in Sec.~\ref{OQC}, an additional two-qubit entangling gate is needed to achieve universality. For this, there are various approaches \cite{Knill2001, Nielsen2004, Browne2005, Ewert2019}, and already several experimental demonstrations too \cite{Obrien2003, Kiesel2005, Okamoto2005, Langford2005, Gasparoni2004, Zhao2005, Huang2004, CNOT}, which typically, however, are either destructive or heralded and probabilistic.

On the other hand, the two-mode interaction for a controlled-$Z$ gate can be decomposed as
\begin{align}
	e^{i\pi\hat n_1\hat n_2}&=e^{i\sqrt{\pi}\hat x_a\hat n_1}e^{-i\sqrt{\pi}\hat p_a\hat n_2}e^{-i\sqrt{\pi}\hat x_a\hat n_1}e^{i\sqrt{\pi}\hat p_a\hat n_2},\label{CZdecomp}
\end{align}
with the subscript $a$ denoting an extra ancilla mode.
Combining the approximated controlled phase rotation gate $S_{\bm{\lambda}}(t)$ with a displacement gate,
\begin{align}
	\hat M^{(j)}&:=e^{-i\sqrt{\pi}\hat x_a/2}S^{(a, j)}_{\bm{\lambda}}(\sqrt{\pi})\approx e^{i\sqrt{\pi}\hat x_a\hat n_j},
\end{align}
we can thus approximate the controlled-$Z$ gate by simply applying the same gate four times,
\begin{align}
	e^{i\pi\hat n_1\hat n_2}&\approx\hat M^{(1)}\hat F^\dagger_a \hat M^{(2)}\hat F^\dagger_a \hat M^{(1)}\hat F^\dagger_a \hat M^{(2)}\hat F^\dagger_a\,.\label{CZapproximation}
\end{align}
Here the Fourier gate $\hat F_a=\exp(i\frac{\pi}{2}\frac{\hat x_a^2 + \hat p_a^2}{2})$ is performed on the ancilla and the superscripts indicate the different qubits and qumodes the operators act upon.
A simple calculation shows that the four displacements combined just perform the operation $\hat F_1^\dagger\hat F_2^\dagger$ and can thus be replaced by two additional Fourier gates.
In order to evaluate the quality of the approximated controlled-$Z$ gate, we use the worst-case fidelity after tracing out the ancillary mode,
\begin{figure}
	\centering
	\includegraphics[width=\columnwidth]{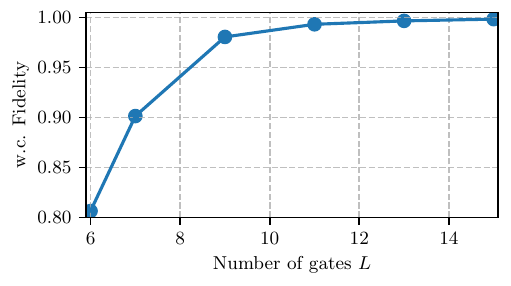}
	\caption{Worst-case fidelity $F_{wc}$ of the approximated controlled-$Z$ gate for optimized parameter sets $\bm{\lambda}$ of different length $L$.}
	\label{minimF}
\end{figure}
\begin{align}
	&F_{wc}=\min_{\left|\psi\right>,\left|\phi\right>}F\left(\text{tr}_a\left[\text{CZ}_\approx\left|0\right>_a\left|\psi\right>_1\left|\phi\right>_2\right],\text{CZ}\left|\psi\right>_1\left|\phi\right>_2\right),\nonumber\\
	&\left|\psi\right>,\left|\phi\right>\in\left\{\alpha\left|0\right>+\beta\left|1\right>\big||\alpha|^2+|\beta|^2=1\right\},
\end{align}
with the fidelity of a density matrix and a pure state
\begin{align}
    F(\hat \rho, \left|\chi\right>) = \bra{\chi}\hat \rho \left|\chi\right>.
\end{align}
In Fig.~\ref{minimF} this worst-case fidelity is calculated for optimized parameter sets $\bm{\lambda}$ of different length $L$. The exact number of the different elementary gates and the maximal required cubic phase gate strength can be found in Table~\ref{CZtable}. The explicit parameter sets as well as the numerical calculations can be found in the Appendix.

\begin{table}
	\centering
        \caption{Number of elementary gates, required cubic phase gate strength $r_{max}$ and worst-case fidelity $ F_{wc}$ of the approximated controlled-$Z$ gate given by Eq.~\eqref{CZapproximation} for the optimized parameter sets of different length $L$.}\label{CZtable}
        \begin{ruledtabular}
	\begin{tabular}{lccccc}
        & Cubic phase && Beam & Fourier & Worst-case \\[-0.7mm]
        $L$ & gates & $r_{max}$ & splitters & gates & Fidelity \\[1mm]\hline
	6 & $4\times18$ & 0.069 & $4\times18$ & $4\times6+6$ & 0.806 \\
        7 & $4\times21$ & 0.062 & $4\times21$ & $4\times6+6$ & 0.901 \\
        9 & $4\times27$ & 0.052 & $4\times27$ & $4\times8+6$ & 0.980 \\
        11 & $4\times33$ & 0.044 & $4\times33$ & $4\times10+6$ & 0.993 \\
        13 & $4\times39$ & 0.039 & $4\times39$ & $4\times12+6$ & 0.997 \\
        15 & $4\times45$ & 0.035 & $4\times45$ & $4\times14+6$ & 0.998
	\end{tabular}
        \end{ruledtabular}
\end{table}

The results show that it is possible to achieve a high-fidelity controlled-$Z$ gate with close to ten cubic QND gates. Provided a source of deterministic single-mode cubic phase gates is available, this approximation could supersede the probabilistic controlled-NOT gates and allow for deterministic and universal quantum computing using photonic qubits. Moreover, Eq.~\eqref{CZdecomp} can, with little to no modification, be used to implement a CV self-Kerr as well as cross-Kerr gate when acting on CV states beyond just single-photon states like in the above scheme. While not further investigated in this paper, both have a wide array of applications.

For instance, a unit-fidelity, non-Clifford gate on physical GKP qubits
is possible with a self-Kerr-based interaction $\sim\hat n^2$ \cite{Royer2022}.
The CZ gate as discussed in this section could also be realized near-deterministically based on a cross-Kerr-interaction number-QND approach 
including feedforward \cite{Nemoto2004}, though the single-mode model for treating the naturally available nonlinear optical interactions in this case must be handled with care \cite{Shapiro2006}.
Compared with other schemes that aim at obtaining quartic Kerr-type interactions and either employ large numbers of elementary cubic or quartic gates, starting in the thousands for similar fidelities \cite{LLoyd1999, Sefi2011, Douce2019}, or a smaller number of then absolutely necessary quartic gates \cite{Sefi2013},
our scheme only requires about a hundred elementary cubic single-mode gates. This is also distinct from the approach of Ref.~\cite{Johnsson21} where a CV single-mode self-Kerr gate is obtained conditionally with Gaussian ancilla states and operations including homodyne measurements,
assuming a two-mode controlled phase rotation gate is available, e.g., from Faraday interactions in atomic ensembles.
Complementary to this, our paper demonstrates how to optically get the controlled phase rotation gate in the first place and then how to combine four such gates, instead of just a single one as in \cite{Johnsson21}, to obtain CV Kerr gates in an unconditional and measurement-free fashion. Nonetheless, our decompositions for controlled phase rotation gates could also be applied to the scheme of Ref.~\cite{Johnsson21}.

Quantum error correction based on a system of sufficiently many physical, photonic qubits, for which the entangling gate discussed in this section works, would nevertheless come with a large experimental overhead.
A more hardware-efficient, and hence potentially promising approach,
makes use of bosonic quantum error correction codes with physical oscillator states 
clearly beyond average photon numbers of one.
An important candidate for this is the GKP code, which we treat next.

\subsection{GKP state generation}

In general, quantum error correction on a system of qubits typically comes with a large computational and hence experimental overhead. In this context, an interesting approach towards an optical fault-tolerant quantum computer employs ``brighter'' optical oscillator states rather than optical dual-rail qubits to encode the logical qubits. The probably most prominent of these encodings is the GKP code presented by Gottesman, Kitaev, and Preskill in 2001 \cite{GKP}, which already includes the possibility to correct errors that become manifest -- and can be formally expanded -- as small shifts in phase space. While experimental demonstrations of GKP-type states have been generally out of reach for many years and only happened very recently in the circuit-QED \cite{Campagne2020} and ion-trap \cite{Fluehmann2019} platforms, an optical, photonics-based realization of these highly non-Gaussian, nonclassical states appears very challenging \cite{KonnoExperimentalGKP}, though some theoretical proposals exist (see, e.g., Refs.~\cite{Eaton2022, Fukui2022, Quesada2019, Su2019, Vasconcelos2010}), some of which depending on Kerr-type ``elementary'' or other quartic gates \cite{Quesada2019, Pirandola2004, FukuiEndo2022}. Our schemes are solely based upon single-mode cubic gates and passive linear optics, unlike, for instance, the recent proposal of Ref.~\cite{Yanagimoto2022} that requires suitable quantum optical multimode Hamiltonians of cubic order. 

In this section, we give a brief overview of GKP states, codes, and gates, before we apply our optical gate decomposition method to a measurement-based \cite{GKP} and a measurement-free \cite{Hmf} GKP state generation method. We shall also consider the creation of arbitrary logical GKP states including the so-called magic states.   

\subsubsection{GKP states, codes, and gates}\label{GKPSGscg}

\begin{figure*}
    \centering
    \includegraphics{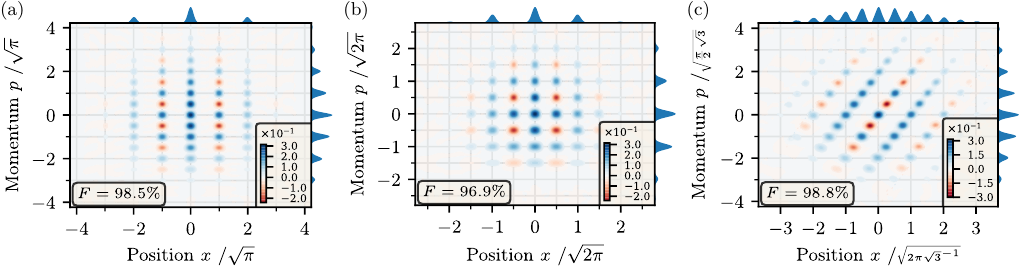}
    \caption{Wigner functions of different approximated GKP states. (a) Approximated $\left|0\right>_{\text{GKP}}$ state for $d=2\sqrt{\pi}$, 12.2 dB squeezing $(n=5)$, and $L=7$ cubic QND gates. (b) Approximated qunaught GKP state for $d=\sqrt{2\pi}$, 11.5 dB squeezing $(n=2)$, and $L=9$ cubic QND gates. (c) Approximated hexagonal GKP state for $d=\left(2\pi\sqrt{3}\right)^{1/2}$, 11.9 dB squeezing $(n=4)$, $L=9$ cubic QND gates, $\delta=\frac{\pi}{2}$, and a $-30^{\circ}$-phase rotation gate.}
    \label{Wigner}
\end{figure*}

The ideal GKP codewords are given by
\begin{align}
	\ket{0_L}=\sum_{s\in\mathbb{Z}}\ket{x=ds},&&
	\ket{1_L}=\sum_{s\in\mathbb{Z}}\ket{x=d\left(s+\tfrac{1}{2}\right)},
\end{align}
where $\braket{x}{x=x_0}$ is the Dirac delta function $\delta(x-x_0)$ and $d\in \mathbb{R}$. They are unnormalizable and clearly unphysical states. However, they are orthogonal and can easily be distinguished by a homodyne measurement. Moreover, displacement errors in $\hat x$ below a threshold of $\frac{d}{4}$ preserve the logical information. Using the Poisson summation formula we find that
\begin{align}
	\ket{0_L}=\sum_{s\in\mathbb{Z}}\ket{p=\tfrac{2\pi}{d}s},&&
	\ket{1_L}=\sum_{s\in\mathbb{Z}}(-1)^s\ket{p=\tfrac{2\pi}{d}s},
\end{align}
with $\braket{p}{p=p_0}=\delta(p-p_0)$.
Therefore the same holds true for displacement errors in $\hat p$ below a threshold of $\frac{\pi}{d}$.
This is remarkable, as an arbitrary error in a CV system can be expanded in terms of displacements,
\begin{align}
	\mathcal{E}(\hat \rho)=\int_{\mathbb{C}^2}d\beta\:d\beta'\ c(\beta, \beta')\thinspace\hat D(\beta)\hat \rho\hat D^\dagger(\beta'),
\end{align} 
with the displacement operator $\hat D(\alpha) = \exp(\alpha\hat a^\dagger - \alpha^*\hat a)$ \cite{GKP}.
The ideal GKP code can thus correct any error with sufficiently small support of $c(\beta, \beta')$.
Choosing $d=2\sqrt{\pi}$ so that $\frac{\pi}{d}=\frac{d}{4}$ is referred to as square GKP code.
The logical gates of the DV universal gate set of Eq.~\eqref{DVUgateset} for the square code are given by
\begin{align}
	\hat H_L=e^{i\frac{\pi}{2}\frac{\hat x^2 + \hat p^2}{2}},&&
	\hat S_L=e^{i\frac{\hat x^2}{2}},&&
	\text{CNOT}_L=e^{-i\hat x_1\hat p_2}.
\end{align}
Moreover, the Pauli operators $\hat X_{L}$, $\hat Y_{L}$, and $\hat Z_{L}$ can be straightforwardly realized using displacements, especially 
$\hat X_L=e^{-i\sqrt{\pi}\hat p}$ and $\hat Z_L=e^{i\sqrt{\pi}\hat x}$. The stabilizers of the code can then be written as $\hat X_L^2$ and $\hat Z_L^2$.
The syndrome measurements for displacement errors in $\hat x$ and $\hat p$ can be done subsequently by linearly coupling the GKP-encoded signal qubit with a suitable GKP ancilla qubit followed by corresponding homodyne detections. 
Notably, provided GKP states are available, all these operations are Gaussian and thus comparably easy to realize experimentally. This is one of the main advantages of the GKP code.

One gate from the DV universal gate set that still has not been considered yet is the non-Clifford gate $\hat T$. In the original proposal, it was suggested to employ the relation
\begin{align}
	\hat T_L=\exp\left(\tfrac{i\pi}{4}\left(2\left(\tfrac{\hat x}{\sqrt{\pi}}\right)^3+\left(\tfrac{\hat x}{\sqrt{\pi}}\right)^2-2\left(\tfrac{\hat x}{\sqrt{\pi}}\right)\right)\right),\label{wrongMagicGate}
\end{align}
which holds true for the ideal GKP states. However, it was recently demonstrated that this operator is unsuitable for approximate finite-energy GKP states \cite{Hcpg}. Instead, a ``magic gate'' can be obtained by ``magic state injection'', a technique similar to gate teleportation using a logical magic state $\ket{T_L}=\tfrac{\ket{0_L}+e^{i\frac{\pi}{4}}\ket{1_L}}{\sqrt{2}}$ as off-line resource \cite{Konno2021magic}.
In general, it is possible to choose $d\neq2\sqrt{\pi}$. For example, the so-called ``qunaught state'' $\ket{q}_\text{GKP}$ with $d=\sqrt{2\pi}$ fulfils $\hat F\ket{q}_\text{GKP}=\ket{q}_\text{GKP}$ and is useful when entangling two GKP states using a beam splitter to obtain a GKP Bell state \cite{Walshe2020}. Another common choice is the hexagonal GKP code with $d=2\left(2\pi/\sqrt{3}\right)^{\frac{1}{2}}$ and $\hat H_L =\hat F\left(\tfrac{\pi}{3}\right)$, which can correct arbitrary displacements below the threshold $(\tfrac{\pi}{2\sqrt{3}})^{\frac{1}{2}}$, related to the closest packing of circles in two dimensions. The Wigner functions of different GKP states are shown in Fig.~\ref{Wigner}.

The most common and practical approach to obtain physical approximations of the ideal GKP states is to replace the Dirac delta functions by Gaussian curves of width $k^{-1}$ and introduce an overall Gaussian envelope of width $k$. Therefore, the resulting states
\begin{align}
	\left<x|0\right>_{\text{GKP}}(k, d)\propto \sum_{s\in\mathbb{Z}} \exp\left(-\frac{(ds)^2}{2k^2}-\frac{k^2(x-d s)^2}{2}\right)\label{GaussianGKP}
\end{align}
can be referred to as Gaussian GKP states, despite being highly non-Gaussian.
In the case of high squeezing, where $k$ is large, these Gaussian GKP states approach the ideal codewords and the probability to misidentify the two nonorthogonal states $\ket{0}_{\text{GKP}}$ and $\ket{1}_{\text{GKP}}$ becomes exponentially small,
\begin{align}
	P_\text{Error}<\frac{2}{\pi k}e^{-\tfrac{\pi}{4}k^2}.
\end{align}
Moreover, they fulfill the relation
\begin{align}
	\ket{j}_\text{GKP}\propto\int_{\mathbb{C}}d\alpha\ \exp\left(-|\alpha|^2k^2\right)\hat D(\alpha)\ket{j_L},
\end{align}
with $j=0, 1$ \cite{GKPEquivalence}. Hence, the Gaussian GKP states can be regarded as ideal GKP states that have undergone a coherent displacement error of Gaussian distribution. This implies that the presented operators do approach their corresponding logical gates for large $k$, as Gaussian gates act only linearly on the operators $\hat x$ and $\hat p$. At the same time, this illustrates why the nonlinear operation defined in Eq.~\eqref{wrongMagicGate} does not converge to the gate $\hat T_L$ even for high squeezing.

The process of error correction also stays unchanged for approximate GKP states, as the code was designed to correct small displacements. Although the intrinsic error of the approximation reduces the margin of external errors that can be successfully corrected, simulations show that even for relatively low squeezing, the GKP code outperforms other CV error correction codes when considering photon loss as source of error \cite{GKPPerformance}.

\subsubsection{GKP state generation with conditional measurement}

\begin{figure*}
    \centering \includegraphics{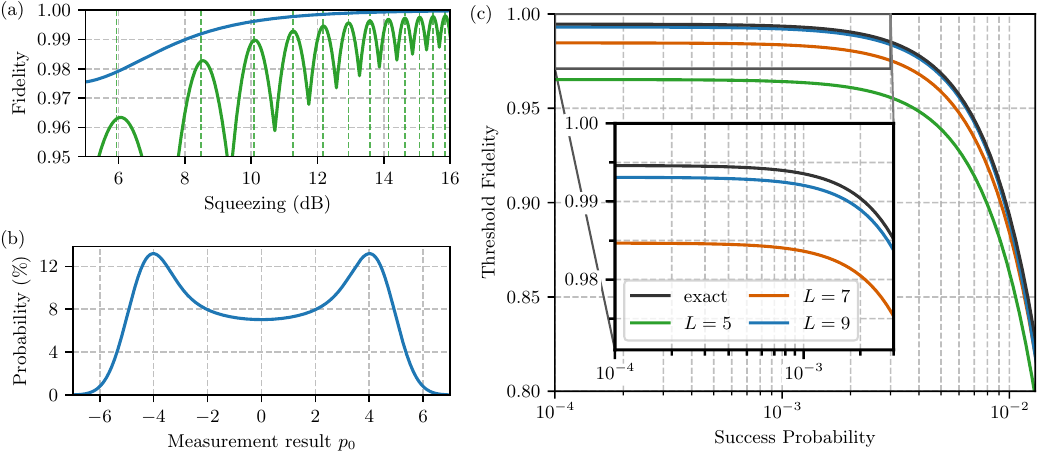}
    \caption{Performance of the approximation for $d=2\sqrt{\pi}$. (a) Fidelity to the target state $\left|0\right>_{\text{GKP}}$ of the absolute value (blue) and after the phase correction (green) given $p_0=0$. The vertical green lines indicate the different $n$'s of Eq. \eqref{quantizedk}. (b) Probability distribution of the homodyne measurement for a squeezing of 12.2 dB $(n=5)$. (c) Threshold fidelity against the success probability for different $p_0$-conditionings $|p_0|\leq \text{const.}$ for a squeezing of 12.2 dB $(n=5)$ and different approximations to the controlled phase rotation gate. $L$ represents the number of cubic QND gates used.}
    \label{CorrectionsABC}
\end{figure*}

While basic operations on the GKP code space as well as error correction are easily implementable in an optical context, the creation of high-quality optical GKP codewords still has not been accomplished 20 years after the original GKP proposal \footnote{However, note that there is a very recent quantum optical experiment demonstrating the creation of photonic states with GKP-type features \cite{KonnoExperimentalGKP}}. However, a creation scheme as old as the code itself can also be found in the original Ref.~\cite{GKP}: Let a controlled phase rotation gate act upon a squeezed vacuum together with a ``meter'' and then measure the meter's phase. More recently, in Ref.~\cite{Weigand2020}, this idea was revisited with a focus on an implementation in the circuit-QED platform.
We shall discuss this particular concept for GKP state generation
in a little more detail.

Here, we intend to propose a slightly altered version, tailored to work in a purely optical setting, based on the presented decompositions. First, we start with a squeezed vacuum together with a displaced state as a meter. Second, we replace the controlled phase rotation gate with our approximation to obtain
\begin{align}
	\left|\psi_{k, \alpha}(t)\right>:=&S_{\bm{\lambda}}(t)\left(e^{-i\frac{\ln(k)}{2}(\hat x\hat p+\hat p\hat x)}\left|\text{vac}\right>\right)_1\left(e^{-i\alpha\hat p}\left|\text{vac}\right>\right)_2.
\end{align}
Third, instead of phase estimation \cite{Weigand2020}, we use a simple homodyne measurement in the quadrature $\hat p_2$ whilst conditioning the measurement result to $p_0\approx 0$. In the ideal case of an infinite displacement and exactly $p_0=0$, this measurement will then fix the strength of the performed controlled phase rotation to multiples of $\pi$ resulting in periodic peaks in the position wave function of mode 1 with a Gaussian envelope given by its initial squeezing. The case of a finite displacement and general measurement outcomes $p_0$ can also be treated analytically.
Regarding the equations
\begin{align}
	0=&S_{\bm{\lambda}}(t)(\sqrt{2}\hat a_2-\alpha) S_{\bm{\lambda}}(-t)\left|\psi_{k, \alpha}(t)\right>,\\
	0=&S_{\bm{\lambda}}(t)(k^{-2}\hat x_1+i\hat p_1) S_{\bm{\lambda}}(-t)\left|\psi_{k, \alpha}(t)\right>
\end{align}
in momentum and position space, replacing $\hat x_1$ with $i\partial_{p_1}$ and $\hat p_2$ with $-i\partial_{x_2}$, respectively, gives us two differential equations with the normalized solution
\begin{align}\begin{split}
		\left<x_1, p_0|\psi_{k, \alpha}(t)\right>=&\frac{N}{\sqrt{A}}\thinspace\exp\left({\frac{-x_1^2}{2k^2}-\frac{B}{2A} p_0^2-i\frac{\alpha}{A} p_0}\right)\\
		&\times\exp\left(\frac{\alpha^2P_{xx}[tx_1]}{2A}\right),\label{GKPstate}
	\end{split}\\
	N=&\frac{1}{\sqrt{\pi k}}\thinspace\exp\left(i\varphi-\frac{\alpha_R^2}{2}\right),
\end{align}
with $A=P_{xx}[tx_1]+iP_{px}[tx_1]$, $B=P_{pp}[tx_1]-iP_{xp}[tx_1]$, the real part $\alpha_R$ and a global phase $\varphi$.
Here the notation $\sqrt{A}$ is used for the solution to the differential equation
\begin{align}
	\frac{f'(x)}{f(x)}=\frac{1}{2}\frac{A'(x)}{A(x)}.
\end{align}
The imaginary phase of the function $\sqrt{A}(x)$ thus covers the full range of $(-\pi, \pi]$ instead of the common $(-\frac{\pi}{2}, \frac{\pi}{2}]$.

Before we look at the approximation's impact on the scheme, let us consider the case of an ideal controlled phase rotation gate. Replacing the polynomials with sine and cosine, respectively, we get
\begin{align}\begin{split}
		\left<x_1, p_0|\psi_{k, \alpha}(t)\right>&\propto\exp\left({-\frac{x_1^2}{2k^2}-\frac{\left|\alpha\right|^2}{2}\sin^2(tx_1+\delta)}\right)\\
		&\times\exp\left(i\frac{tx_1}{2}+i\frac{\left|\alpha\right|^2}{4}\sin(2tx_1+2\delta)\right)\\
		&\times\exp\left(-ip_0\left|\alpha\right|\exp(itx_1+i\delta)\right), \label{terms}
	\end{split}
\end{align} 
with $\alpha=\left|\alpha\right|\,e^{i\delta}$. In order to obtain the correct spacing and squeezing of the GKP code, we must have $t = \pi/d$ and $\left|\alpha\right| = k/t$. The initial phase of the displaced state $\delta$ provides an elegant way of shifting the peaks, while keeping the Gaussian envelope centered. For the state $\left|0\right>_{\text{GKP}}$, however, we will be setting it to $\delta=0$.

\begin{table*}
	\centering
        \caption{List of resources needed for the different presented GKP state generation schemes. The maximal required cubic phase gate strength of each scheme is denoted by $r_{max}$.}\label{GKPtable}
        \begin{ruledtabular}
	\begin{tabular}{lccccccccc}
        & Target && Squeezing & Cubic phase && Beam & Fourier & Displace- &\\[-0.7mm]
        Figure & state & Fidelity & [dB] & gates & $r_{max}$ & splitters & gates & ments & Other \\[1mm]\hline
	\ref{Schema} & $\left|0\right>_\text{GKP}$ & 0.909 & 12.2 & 7 & 0.130 & 9 & 2 & 2 &  \\
        \ref{Wigner} (a) & $\left|0\right>_\text{GKP}$ & 0.985 & 12.2 & 15 & 0.068 & 21 & 6 & 2 &  \\
        \ref{Wigner} (b) & $\left|q\right>_\text{GKP}$ & 0.969 & 11.5 & 19 & 0.066 & 27 & 8 & 2 &  \\
        \ref{Wigner} (c) & hex. $\left|0\right>_\text{GKP}$ & 0.988 & 11.9 & 19 & 0.100 & 27 & 8 & 2 & $e^{-i\frac{\pi}{6}\hat n}$ \\
        \ref{magicState} (b) & $\left|T\right>_\text{GKP}$ & 0.927 & 12.0 & 33 & 0.036 & 48 & 16 & 2 &  \\
        \ref{Hastrup_GKP} & $\left|1\right>_\text{GKP}$ & 0.972 & 11.5 & $4\times10\times24$ & 0.061 & $4\times10\times54$ & $4\times10\times12+8$ &  & $\left|0\right>_\text{DR}$ \\
        \ref{WingnerHmagic} & $\left|T\right>_\text{GKP}$ & 0.991 & 11.7 & $19+2\times44$ & 0.153 & $27+2\times99$ & $8+2\times22+4$ & 3 & $\left|T\right>_\text{DR}$
	\end{tabular}
        \end{ruledtabular}
\end{table*}

With these parameters set, let us look at the terms of Eq.~\eqref{terms} line by line. The first line gives the absolute value of the waveform and approximates $\left<x|0\right>_{\text{GKP}}$ quite well for $k\gg1$, since
\begin{align}
	\exp\left({-\frac{k^2d^2}{2\pi^2}\sin^2\left(\frac{\pi x}{d}\right)}\right)\stackrel{k\gg1}{\approx} \sum_{s\in\mathbb{Z}} \exp\left(-\frac{k^2(x-d s)^2}{2}\right).
\end{align}
The second line gives the phase of the waveform. On the one hand, the first term is due to the negligence of the $-\frac{1}{2}$ term in $\hat n= \hat x^2 + \hat p^2-\frac{1}{2}$ and easily corrected by introducing a corrective displacement of $t/2$. On the other hand, the second term arises from the homodyne measurement and needs a bit more attention. As it takes the same form for every peak, it is sufficient to regard its impact on the fidelity of two single Gaussians,
\begin{align}\begin{split}
	&\left|\int e^{-k^2\varepsilon^2} \exp\left({i\frac{k^2d^2}{4\pi^2}\sin\left(\frac{2\pi \varepsilon}{d}\right) - i \varepsilon\cdot c}\right)d\varepsilon\right|\\
	\approx&\int e^{-\varepsilon^2} \cos\left(\frac{c'}{k}\varepsilon-\frac{\pi }{3kd}\varepsilon^3+\mathcal{O}\left(\frac{1}{k^{3}}\right)\right)\frac{d\varepsilon}{k}\\
	=&\int e^{-\varepsilon^2} \left(-\frac{c'^2}{k^2}\frac{\varepsilon^2}{2}+\frac{c'}{k^2} \frac{\pi\varepsilon^4}{3d}+\mathcal{O}\left(\frac{1}{k^{4}}\right)\right)\frac{d\varepsilon}{k} + \text{const.}\\
	=&\frac{\sqrt{\pi}}{k^3}\left(-\frac{c'^2}{4}+c'\frac{\pi}{4d} + \mathcal{O}\left(\frac{1}{k^{2}}\right)\right) + \text{const.}\,,
\end{split}\end{align} with an additional displacement $c$ and $c'=-c+k^2/2t$. Consequently, given $k\gg1$, the approximation is best for $c=k^2/2t-t/2$ and a total corrective displacement of $k^2/2t$. On the other hand, in order to take the same form for every peak, the additional displacement must fulfill $c=2t\cdot n$ with $n\in\mathbb{Z}$. This leads us to the following condition for the optimal squeezing parameter:
\begin{align}
	k=2t\sqrt{n+1/4},\quad n\in\mathbb{Z}\,.
	\label{quantizedk}
\end{align}
When states with different spacing $t$ but identical squeezing $k$ are needed, it is useful to choose a squeezing parameter for which both states show minimal deviation from the optima, for example $2\sqrt{n+1/4+1/20}) = \sqrt{n' + 1/4-1/20}$. The performance of this phase correction compared to that of the absolute value can be seen in Fig.~\ref{CorrectionsABC}(a). A change of the fixed peak spacing $d$ whilst keeping $n$ constant is found to have a negligible effect on the calculated fidelities.

The third line gives the error introduced by a measurement result of $p_0\neq0$. Although this term will later prove useful in the creation a GKP magic state, here it is nothing but an uncorrectable but preventable error. Accepting a larger interval of measurement results $p_0$ increases the maximum error, while a smaller acceptance interval lowers the success probability. The introduced errors and success probabilities for a squeezing of 12.2 dB $(n=5)$ can be found in Fig.~\ref{CorrectionsABC}(c). Larger $n$, whilst leading to an increased maximum fidelity, are found to be accompanied by smaller success probabilities.

Up to now we have not considered the impact of an approximated controlled phase rotation gate yet. In order to do so, we optimize parameter sets $\bm{\lambda}$ of different length $L$ to maximize the fidelity of the approximation and its target state for $p_0=0$. The results are plotted in Fig.~\ref{CorrectionsABC}(c). They show that it is possible to achieve a practically perfect approximation to the controlled phase rotation gate with less than ten as well as fidelities over 96\% with merely five cubic QND gates. As a general rule, the more peaks the target state has, the more gates are needed to approximate it properly. Hence, states such as $+$, $-$, qunaught, and hexagonal GKP states tend to need a higher squeezing as well as a higher operator count to achieve the same fidelities as the $\left|0\right>_{\text{GKP}}$ state. The parameters needed to obtain the different GKP states and encodings are summarized in Table~\ref{table}, while the resulting Wigner functions of the GKP states considered here, given similar squeezing and operator counts, are shown in Fig.~\ref{Wigner}.

Note that all these GKP states can be created using $2L+1$ cubic phase gates of the same gate strength. This is demonstrated in the exemplary circuit of Fig.~\ref{Schema} and is likely to be significant for their experimental feasibility. In fact, a recent experimental demonstration \cite{Fcpgexp} would correspond to a cubic-phase-gate strength parameter $r$ in Eq.~(\ref{CPGHeisenberg}) as $r \approx 0.17$. In Appendix~\ref{AppOPS} we present optimized parameter sets for the scheme of Fig.~\ref{Schema} where we can choose all gate strengths identical and below a value of 0.17. Weaker gate strengths may allow to improve the final state fidelities for larger gate concatenations. When sticking to the scheme of Fig.~\ref{Schema} the optimized nonidentical gate strengths include a maximal value of $~0.13$ (see Appendix~\ref{AppOPS}). For an easy comparison the amount of elementary gates as well as the maximal cubic-phase-gate strength $r_{max}$ needed to generate the different GKP states are also summarized in Table~\ref{GKPtable}.

\subsubsection{Creating a GKP magic state}

\begin{figure*}
    \centering
    \includegraphics{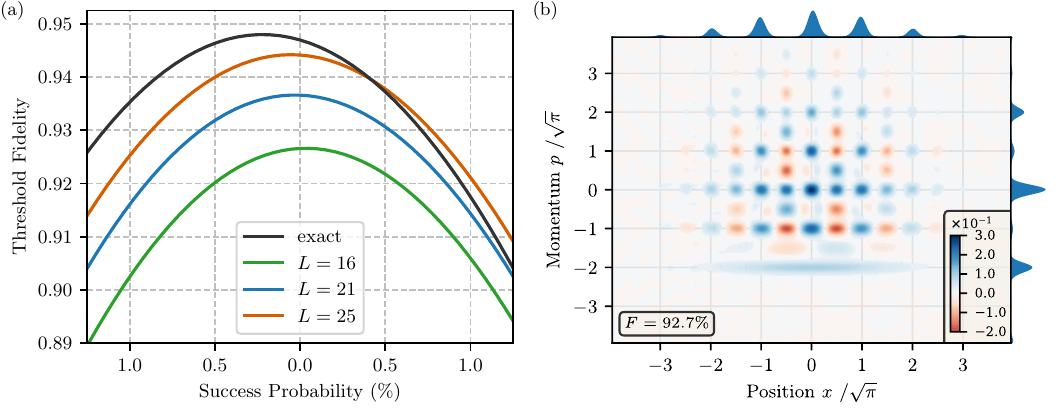}
    \caption{Approximating a GKP magic state. (a) Fidelity to the target state $\left|T\right>_{\text{GKP}}$ plotted against the success probability for a squeezing of 12.0 dB $(n=1)$ and different approximations to the controlled phase rotation gate. $L$ represents the number of cubic QND gates employed. As the fidelity is not symmetric around $p_0 = \frac{\pi^2}{8kd}$, success is defined one-sided: $\text{const.}\leq p_0\leq \frac{\pi^2}{8kd}$ to the left and $\frac{\pi^2}{8kd}\leq p_0\leq \text{const.}$ to the right. (b) Approximated $\left|T\right>_{\text{GKP}}$ state using $d=\sqrt{\pi}$, $p_0=\frac{\pi^2}{8kd}$, 12.0 dB squeezing $(n=1)$ and $L=16$.}
    \label{magicState}
\end{figure*}

Besides creating the logical GKP codewords, there is one more obstacle to overcome in order to achieve full universality with optical GKP qubits: a qubit non-Clifford gate such as $\hat T$ that acts logically in the corresponding way when applied upon GKP states, like $\hat T_L$ of Eq.~(\ref{wrongMagicGate}) as originally proposed in Ref.~\cite{GKP} and discussed in Secs.~\ref{OQCuqc} and \ref{GKPSGscg}.
The original approach though 
performs rather poorly on finitely squeezed states \cite{Hcpg} and an optimized version was found to saturate at a logical fidelity of about 95\% for high squeezing \cite{Konno2021magic}. The more promising approach is therefore the gate teleportation of a magic state $\left|0\right>_\text{GKP}+e^{i\frac{\pi}{4}}\left|1\right>_\text{GKP}$ \cite{Konno2021magic, GKP},
as was also mentioned before. 
Here we want to discuss how our optical GKP creation scheme can be used to approximate a GKP magic state. The underlying idea is to change the $p_0$ conditioning and make use of the introduced error.

Let us take a look at the $p_0$ dependence of Eq.~\eqref{terms},
\begin{align}
	\exp\left(-ip_0\frac{kd}{\pi}\exp(itx_1)\right).\label{p0Error}
\end{align}
Accepting the measurement result only when $p_0\approx \frac{\pi^2}{8kd}$ we obtain a phase of
\begin{align}
	\exp\left(i\frac{\pi}{4}\frac{1-\cos(tx_1)}{2}\right),
\end{align}
up to a global phase. This is a fairly good approximation of a logical $\hat T$ gate in the sense that every second GKP peak obtains a phase of $e^{i\frac{\pi}{4}}$.
\begin{table}
	\centering
        \caption{Parameters used to approximate GKP states of different encodings.}\label{table}
        \begin{ruledtabular}
	\begin{tabular}{llccc}
		Encoding & State & $t$ & $p_0$ & $\delta$ \\\hline
		Square & $\left|0\right>_\text{GKP}$ & $\sqrt{\pi}/2$ & 0 & 0\\
		Square & $\left|1\right>_\text{GKP}$ & $\sqrt{\pi}/2$ & 0 & $\frac{\pi}{2}$\\
		Qunaught & $\left|q\right>_\text{GKP}$ & $\sqrt{\pi/2}$ & 0 & 0\\
		Hexagonal & $\left|0\right>_\text{GKP}$ & $\left(\pi/(2\sqrt{3})\right)^{1/2}$ & 0 & $\frac{\pi}{2}$\\
		Square & $\left|T\right>_\text{GKP}$ & $\sqrt{\pi}$ & $\frac{\pi}{8|\alpha|}$ & 0
	\end{tabular}
        \end{ruledtabular}
\end{table}
This means, however, that in order to get a magic state, we need to start with a peak spacing of $d=\sqrt{\pi}$ provided with a $\left|+\right>_{\text{GKP}}$ state. When using the same squeezing as for the $\left|0\right>_{\text{GKP}}$ state, this implies that the fidelity of the magic state will be inherently worse. The resulting fidelity saturates at 94.8\% for a comparable squeezing of 12.0 dB $(n=1)$ independent of the operator count $L$, but with a higher success probability (see Fig.~\ref{magicState}). In order to obtain higher fidelities, a higher squeezing must be chosen, for example, a squeezing of 14.5 dB $(n=2)$ allows for fidelities up to 97.7\%. It might also be possible to use the more accurate logical GKP states in order to distil magic states of higher fidelity. On the other hand, the scheme presented in the next section can be used to create a magic state from a high-fidelity $\left|0\right>_{\text{GKP}}$ state.
Let us finally mention that when using $\left|+\right>_{\text{GKP}}$ states to initiate the logical qubits, logical qubits and magic states could be generated in parallel given a two-part acceptance interval
for the measurement outcomes.

\subsubsection{Measurement-free GKP state generation}

One of the main advantages of a purely optical quantum computer is the achievable clock rate. Therefore it would be desirable to have a deterministic, measurement-free scheme as opposed to the probabilistic GKP state creation of the preceding sections.
In Ref.~\cite{Hmf}, the authors present a measurement-free GKP creation scheme using the Rabi-type Hamiltonian gates $e^{i\hat p\hat \sigma_x}$ and $e^{i\hat x\hat \sigma_y}$, which are readily available in trapped-ion and superconducting circuit platforms, where the spin operators can either act upon a real or an ``artificial'' two-level atom. In fact, there are already experimental demonstrations in these platforms based on similar concepts \cite{Neeve2022, Eickbusch2022}. Here, however, we shall apply our decomposition results for purely optical Rabi-type Hamiltonians of Sec.~\ref{Rabidecompos} to the GKP state generation scheme of Ref.~\cite{Hmf}.

Using the gates $U_k=e^{iu_k\hat x\hat \sigma_y}$, $V_k=e^{iv_k\hat p\hat \sigma_x}$, and $W_k=e^{iw_k\hat x\hat \sigma_y}$ and applying them on an infinitely squeezed state $\left|x_0\right>$ and a qubit ancilla state $\alpha\left|+\right>_a+\beta \left|-\right>_a$ as an input, one obtains 
\begin{align}
	&W_kV_kU_k\left|x_0\right>\left(\alpha\left|+\right>_a+\beta \left|-\right>_a\right) \nonumber\\
	=&W_kV_k\left|x_0\right>\left(A\left|+\right>_a+B\left|-\right>_a\right) \nonumber\\
	=&W_k\left(A\left|x_0-v_k\right>\left|+\right>_a+B\left|x_0+v_k\right>\left|-\right>_a\right)\label{HastrupProtocol}\\
	=&\cos(w_kx_0)\left(A\left|x_0-v_k\right>\mp B\left|x_0+v_k\right>\right)\left\{^{\left|1\right>_a}_{\left|0\right>_a}\right\} \nonumber\\
	+&\sin(w_kx_0)\left(\pm A\left|x_0-v_k\right>-B\left|x_0+v_k\right>\right)\left\{^{\left|0\right>_a\text{, if $v_kw_k=+\frac{\pi}{4}$}}_{\left|1\right>_a\text{, if $v_kw_k=-\frac{\pi}{4}$}}\right. \nonumber
\end{align}with
\begin{align}
	A=\alpha\thinspace \cos(u_kx_0) - \beta\thinspace \sin(u_kx_0),\\
	B=\beta\thinspace \cos(u_kx_0) + \alpha\thinspace \sin(u_kx_0).
\end{align}
As we can see, the displacement gate $V_k$ displaces $\left|x_0\right>$ depending on the state of the qubit, effectively splitting it in two. The disentangling gate $W_k$ then disentangles qubit and qumode again if and only if $v_kw_k=\pm\frac{\pi}{4}$ and $w_kx_0=m\cdot \frac{\pi}{2}$ with $m\in\mathbb{Z}$. The preparation gate $U_k$, on the other hand, rotates the ancilla qubit depending on the position $x_0$ and together with the original qubit state determines the amplitudes of the two resulting squeezed states $\left|x_0-v_k\right>$ and $\left|x_0+v_k\right>$.

When repeating this procedure $N$ times it is thus possible to split the original $\left|x_0\right>$ state into a sum of $2^N$ distinct position states.
\begin{figure}
	\centering
	\includegraphics[width=\columnwidth]{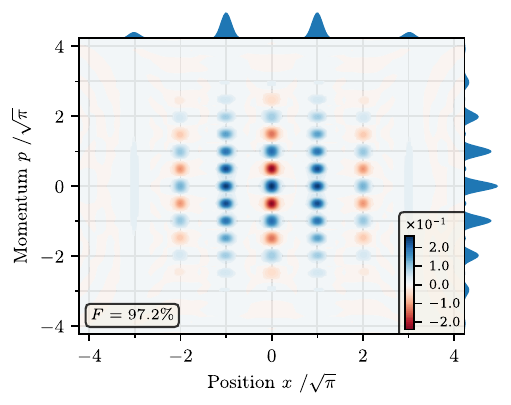}
	\caption{Approximating a $\left|1\right>_{\text{GKP}}$ state using the measurement-free protocol of Ref.~\cite{Hmf} with $N=2$, a squeezing of 11.5 dB, $u_2=0.093$, and $L=60$ CV Toffoli gates for each of the four approximated Rabi-type Hamiltonians.}
	\label{Hastrup_GKP}
\end{figure}
Starting with $\left|x_0=0\right>\left|0\right>_a$ and choosing the $v_k$ and $w_k$ to be
\begin{align}
	v_k &= \begin{cases}
		-\sqrt{\pi}2^{N-1},&\text{if $k=1$},\\
		+\sqrt{\pi}2^{N-k},&\text{if $k>1$},
	\end{cases}\\
	w_k &= \begin{cases}
		-\frac{\sqrt{\pi}}{4}2^{-(N-k)},&\text{if $k<N$},\\
		+\frac{\sqrt{\pi}}{4},&\text{if $k=N$},
	\end{cases}
\end{align}
we obtain the state
\begin{align}
	\left(\sum_{k=1}^{2^N}c_k\left|(2k - 1 - 2^N)\sqrt{\pi}\right>\right)\left|0\right>_a\approx\left|1\right>_\text{GKP}\left|0\right>_a.
\end{align}
The weights of the different peaks $c_k$ are determined by the strengths of the preparation gates $u_k$. Different sets of the latter optimized for different figures of merit can be found in Ref.~\cite{Hmf}. Note that generally setting $u_1=0$ and combining the operators $W_k$ and $U_{k+1}$ reduces the number of required Hamiltonians to $2N$. When introducing finitely squeezed states as an input, the preparation and disentangling gates are no longer exact. After tracing out the qubit ancilla, this leads to a mixed final state. Measuring the ancilla qubit after each iteration could therefore provide pure final states together with overall higher fidelities. However, this is not necessary, as the measurement-free scheme on its own is already able to produce 
high-fidelity states for simply $N\geq2$.

Replacing the Rabi-type Hamiltonian gates with the presented approximations using the $T_{\bm\lambda}(t)$ gate on an optical qumode together with an optical dual-rail qubit, we are able to translate this protocol into a purely optical setting. The resulting $\left|1\right>_\text{GKP}$ state for $N=2$, a squeezing of 11.5 dB, $u_2=0.093$, and $L=60$ CV Toffoli gates for each of the four approximated Hamiltonians can be seen in Fig.~\ref{Hastrup_GKP}. The results show that it is possible to deterministically create GKP states using only single-mode cubic phase gates, beam splitters, phase rotations, and off-line squeezing together with an off-line single-photon qubit ancilla state. 

However, the number of gates needed to properly approximate this protocol is significantly higher than for the probabilistic one. This can be attributed to two factors.
On the one hand, due to an already computationally expensive fidelity calculation, in this case we refrained from doing a parameter optimization and instead used the third-order set $\bm\lambda=\left(0.397, -0.794, -0.0325, 1.54, 0.636, 0.254\right)^T$, which had proven effective in previous applications.
On the other hand, the relatively high gate strength $v_1=2\sqrt{\pi}$, corresponding to a $d=\frac{\sqrt{\pi}}{2}$ in the probabilistic scheme, is generally harder to approximate, as this increases the number of relevant sine/cosine periods.  

\subsubsection{Creating arbitrary logical GKP states}

Another useful application of the above protocol is the creation of arbitrary logical states. While simple changes in the parameters $u_k$, $v_k$, and $w_k$ are sufficient to obtain states of nonsquare encodings such as rectangular and hexagonal $\left|1\right>_{\text{GKP}}$ states \cite{Hmf}, modifying the initial state of the ancilla qubit enables us to create arbitrary logical states. Choosing $u_1=0$, $v_1=-\frac{\sqrt{\pi}}{2}$, and $w_1=\frac{\sqrt{\pi}}{2}$ together with an additional displacement and a $\left|0\right>_{\text{GKP}}$ state as an input, we have
\begin{align}
	&\exp\left(i\frac{\sqrt{\pi}}{2}\hat p\right)W_1V_1U_1\left(\sum_{k\in\mathbb{Z}}c_k\left|2k\sqrt{\pi}\right>\right)\left(\alpha\left|+\right>_a+\beta \left|-\right>_a\right) \nonumber\\
	=&\sum_{k\in\mathbb{Z}}c_k\thinspace(-1)^k\left(\alpha\left|2k\sqrt{\pi}\right>+\beta\left|(2k-1)\sqrt{\pi}\right>\right)\left|0\right>_a.\label{ArbitraryStateGen}
\end{align}
Alternatively, the same result can be obtained for a $\left|1\right>_{\text{GKP}}$ state as an input by straightforward modifications to the parameters.

Two aspects of the final state still need to be addressed.
First, the term $(-1)^k$ is clearly unwanted and in the original proposal of Ref.~\cite{Hmf}, it gets corrected by a displacement in $p$. This, however, introduces linear error terms. Another option is to run the protocol twice using a $\left|+\right>$ ancilla on the first and an ancilla in the sought after logical state on the second run. Moreover, both GKP creation schemes presented in this paper can naturally compensate for this term: in the probabilistic scheme this is done by choosing $n=n_0+\frac{1}{2}$ with $n_0\in\mathbb{N}$, while in the deterministic protocol it is sufficient to invert the sign of $w_N$.
Second, shaping the overall form of the peaks using $U_1$ mixes the coefficients $\alpha$ and $\beta$, and is thus problematic for some aspired states. Moreover, while using a $\left|0\right>_{\text{GKP}}$ state displaced by $\sqrt{\pi}$ as $\left|1\right>_{\text{GKP}}$ state has a negligible effect on the fault tolerance of the code, it heavily influences the calculated fidelities (as much as 10\% for a squeezing of 11.5 dB).
In order to not distort the results, the fidelities are therefore calculated towards the target state $\alpha\left|0\right>_{\text{GKP}}+\beta\exp\left(i\sqrt{\pi}\hat p\right)\left|0\right>_{\text{GKP}}$ instead of the usual $\alpha\left|0\right>_{\text{GKP}}+\beta\left|1\right>_{\text{GKP}}$.

\begin{figure}
	\centering
	\includegraphics[width=\columnwidth]{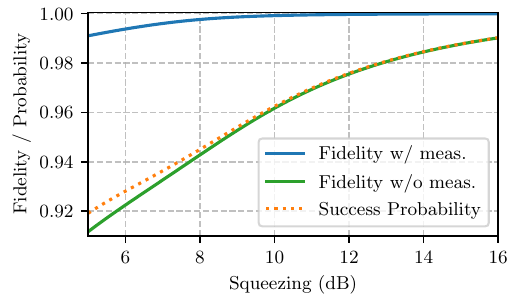}
	\caption{Performance of the protocol of Eq.~\eqref{ArbitraryStateGen} applied on a Gaussian $\left|0\right>_\text{GKP}$ state as in Eq.~\eqref{GaussianGKP} with alternately signed peaks and varying squeezing as an input. The green line gives the fidelity to the magic state $\frac{1}{\sqrt{2}}\left|0\right>_{\text{GKP}}+\frac{(1+i)}{2}e^{i\sqrt{\pi}\hat p}\left|0\right>_{\text{GKP}}$ after tracing over the ancilla qubit. The blue and orange dotted lines, respectively, show the fidelity and success probability when measuring the ancilla qubit.}
	\label{HmagicFidelity}
\end{figure}

In order to test the protocol we are again going to look at the creation of a GKP magic state. Not yet considering the gate approximations, its core performance for different squeezing levels when choosing $\alpha=1/\sqrt{2}$ and $\beta=(1 + i)/2$ is plotted in Fig.~\ref{HmagicFidelity}.
\begin{figure}
	\centering
	\includegraphics[width=\columnwidth]{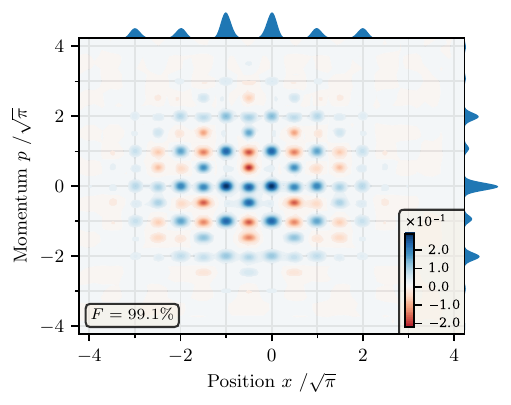}
	\caption{Approximating a GKP magic state using a combination of the two presented creation methods. First, a $\left|0\right>_\text{GKP}$ state with alternately signed peaks is created using the probabilistic scheme with $d=2\sqrt{\pi}$, 11.7 dB squeezing $(n=4.5)$, and $L_1=9$  cubic QND gates. Then the protocol of Eq.~\eqref{ArbitraryStateGen} with $\alpha=1/\sqrt{2}$ and $\beta=(1 + i)/2$ is approximated using $L_2 = 11$ CV Toffoli gates for the two Rabi-type Hamiltonians. Finally, the dual-rail qubit is measured with a success probability of 93.7\%. Without the measurement the same procedure gives a mixed state with a fidelity of 95.3\%.}
	\label{WingnerHmagic}
\end{figure}
The results show that when tracing over the qubit ancilla the fidelity of the resulting mixed state is highly dependent on the squeezing level of the input. Measuring the qubit ancilla, on the other hand, gives a pure state with significantly increased fidelity.

Incorporating the gate approximations of the optical setting, the operators of Eq.~\eqref{ArbitraryStateGen} can be rewritten as 
\begin{align}
	\exp\left(i\frac{\sqrt{\pi}}{2}\hat p\right)\hat F_3T_{\bm{\lambda}}\left(\frac{\sqrt{\pi}}{2}\right)\hat F_3^\dagger\hat F_1^\dagger T_{\bm{\lambda}}\left(\frac{\sqrt{\pi}}{2}\right)\hat F_1.\label{approxHmagic}
\end{align}
Consequently, creating a GKP magic state from a given $\left|0\right>_{\text{GKP}}$ state enables the repeated use of the same approximated Rabi-type Hamiltonian gate $T_{\bm{\lambda}}(t)$ and, as its gate strength is comparably low, allows to approximate the results of Fig.~\ref{HmagicFidelity} well using no more than 10 CV Toffoli gates.
This is in stark contrast to the creation of a $\left|1\right>_{\text{GKP}}$ state from scratch using the same protocol. It is therefore close at hand to combine the probabilistic creation of a high fidelity $\left|0\right>_{\text{GKP}}$ state with Eq.~\eqref{approxHmagic}. The resulting magic state after measuring the ancilla qubit can be seen in Fig.~\ref{WingnerHmagic}. This shows that it is possible to obtain high-fidelity GKP magic states even for relatively low squeezing and a reasonable number of cubic QND as well as CV Toffoli gates. At the same time, it emphasizes the notion that, as opposed to the probabilistic schemes, the deterministic creation of arbitrary logical GKP states, whilst possible, still comes with high requirements on the amount of experimental resources.

\section{Experimental Imperfections: Photon Loss}\label{Loss}

\begin{figure}
    \centering
    \includegraphics{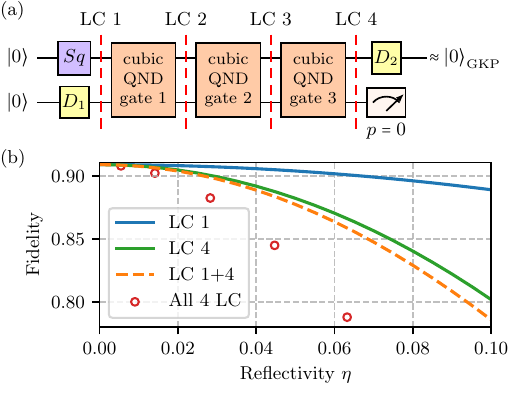}
    \caption{Effects of photon loss on the circuit of Fig.~\ref{Schema}. (a) Positioning of the different photon-loss channels. Each channel consists of two loss beam splitters with a reflectivity of $\eta$, one per mode. (b) Impact on the fidelity of the approximated GKP state for different sets of active photon-loss channels.}
    \label{LossesAB}
\end{figure}

Let us now consider the effect of experimental errors on the presented approximations. In a purely optical setup, these are primarily comprised of photon loss as well as imperfect gates from the universal gate set. As an exhaustive analysis of these errors and their impact on the different schemes goes beyond the scope of this paper, we are going to focus on one exemplary case: the effect of photon loss on the circuit of Fig.~\ref{Schema}. 

Photon loss can be modelled by introducing a beam splitter $\hat B_{1a}(s)$ with the reflectivity $\eta\equiv\sin(s)\ll1$, which reflects a part of the given light mode out into a second mode $\left|\text{vac}\right>_a$ representing the environment. Two of these loss beam splitters are then needed to cover the two modes of the circuit. For simplicity as well as computability, we will not consider losses happening within the cubic QND gates, leaving us with four places in the circuit where the beam splitters can be positioned, namely after each of the three cubic QND gates as well as after the squeezed and displaced input states. This is shown in Fig.~\ref{LossesAB}(a). The impact of these eight loss beam splitters on the fidelity of the resulting GKP state in dependence of the reflectivity $\eta$ can be seen in Fig.~\ref{LossesAB}(b). Besides including all these eight loss beam splitters, the effect of faulty input states and a single erroneous cubic QND gate on their own is also shown. On the one hand, we find that errors of the input states impact the resulting GKP state less than errors occurring later on in the circuit. On the other hand, even for photon loss taking place throughout the scheme the fidelity of the final state still converges to its no-loss value surpassing $90\%$ for $\eta<0.015$. Note that the reflectivity parameter that represents the effect of photon loss in our model describes the fraction of the signal mode operator's amplitude, which is subtracted from the signal. The more standard photon loss probability would then be $\eta^2$, corresponding to a photon transmission of $1 - \eta^2$.    

Another observation is that the impact of multiple errors is less than the sum of their individual impacts. Thus several small errors are less harmful than one large one. 
To which degree these findings hold for other sources of error and the different applications remains untested.
It is crucial that the circuits that we derived for various applications are sufficiently short, i.e., contain only a small number of CV cubic gates, unlike previous decomposition schemes that rely upon commutator approximations \cite{LLoyd1999, Sefi2011}
or also higher nonlinear, namely quartic elementary CV gates
\cite{Douce2019, Sefi2013}. 
In particular, the GKP state generations are not fault tolerant, since the initial quantum optical resource states are not protected against lossy or even faulty CV gates by a quantum error correction code -- the encoded, protected GKP state will only be the result of the application of the CV gate sequences.     
Nonetheless, the tolerable loss rates that we obtained above are comparable with the values proposed in other schemes
and error or loss rates of the order $10^{-3}$ or $10^{-4}$ per gate are not an uncommon requirement of fault-tolerant schemes that are already based on concatenated quantum error correction codes. These thresholds may be reduced, but the necessary codes may then be more complicated, which would in turn have a negative impact on potential experiments.  

Note that for some of the other applications that are possible with our decompositions, loss or fault tolerance can be provided. An example for this would be a non-Clifford gate on GKP qubits that approaches unit fidelity with high-quality GKP states based on a self-Kerr interaction $\sim \hat n^2$ \cite{Royer2022}. 
In this case, the states are protected to some extent against the effect of physical CV error channels, also when sequences of CV gates are applied, provided the GKP states are of sufficient quality
\cite{Menicucci2014, Fukui2017, Fukui2018}.
Nevertheless, also for this application, to combine fault tolerance with scalability, even when a time-domain approach is employed
\cite{FurusawaTimeDomain, AndersenTimeDomain},
it is useful to minimize the length of the CV circuits.
In previous schemes, decompositions for Kerr-type gates would be typically based on long gate sequences involving commutator approximations \cite{LLoyd1999, Sefi2011, Douce2019} or at least quartic elementary CV gates \cite{Sefi2013}.
Generally, the results give confidence that the approximations presented in this paper can function even within the inherently noisy experimental setting, provided an appropriate threshold.

\section{Conclusions}\label{Concl}

In conclusion, we have demonstrated that a limited number of single-mode CV cubic phase gates -- the canonical non-Gaussian gate of standard CV quantum computation --
is a useful and sufficient resource for various elements in photonic quantum computation. This includes optical schemes for DV and CV quantum information processing.
Unlike existing schemes that typically rely on complicated commutator approximations and hence require long CV gate sequences, our decomposition method is ``hybrid'': it employs exact decompositions whilst possible and only reverts back to efficient approximations for mixing the $x$ and $p$ variables.

We have explicitly analyzed the performance of our method for three applications:
(i) a two-qubit two-photon, entangling controlled-$Z$ gate, (ii) a homodyne-measurement-based, conditional, optical GKP state generation scheme based on Gaussian resource states, and (iii) a measurement-free, optical GKP state generation scheme based on Gaussian and single-photon resource states. Our quantitative results imply that tens of single-mode cubic phase gates together with CV Fourier gates and beam splitters
are sufficient to create high-fidelity GKP states, in the form of standard logical Pauli eigenstates or even ``magic states'', and also to realize
a deterministic, two-qubit two-photon controlled-$Z$ gate in an optical setting. The GKP magic states can be employed to obtain a GKP qubit non-Clifford gate by means of gate teleportation techniques \cite{Konno2021magic}. Another option would be to directly apply a quartic self-Kerr gate upon the GKP qubits, which, in principle, would allow for a unit-fidelity GKP non-Clifford gate operation \cite{Royer2022}. We have demonstrated that our efficient decomposition method can be applied to such Kerr gates.
Consequently, single-mode cubic phase gates are a suitable resource for all-optical, fault-tolerant, universal quantum computing.

More specifically, it was shown that the hybrid gate decomposition scheme consisting of exact decomposition techniques and efficient Trotterization is a powerful tool in creating non-Gaussian gates in an optical context. We were able to build a two-mode controlled phase rotation gate $e^{i\hat x_1 \hat n_2}$ as well as the three Rabi-type Hamiltonians corresponding to the three-mode unitaries $e^{i\hat x_1\hat \sigma_{x, S}}$, $e^{i\hat x_1\hat \sigma_{y, S}}$, and $e^{i\hat x_1\hat \sigma_{z, S}}$ where the spin operators refer to the two-mode Schwinger representation. Testing their performance in different applications for photonic qubits, qumodes, as well as a mixture of both, it was shown that in most cases less than 30 single-mode cubic phase gates are already sufficient in providing a good approximation.

Regarding an optical creation of GKP states, this represents a significant improvement over previously known optical generation schemes based on the application of CV circuits that include nonlinear gates.
Compared to a ``Gaussian Boson Sampling'' setup, our presented schemes do not reach the same fidelities to the target state but increase the success probability by multiple orders.
In order to synchronize the optical quantum states and make them available when needed -- quasi on demand, the effect of the calculated success probabilities could be circumvented by the use of all-optical, cavity-based quantum memories \cite{Yoshikawa2013}.
These would even allow to optically store complicated, phase-sensitive, CV states such as GKP states \cite{Hashimoto2019}.

The all-optical quantum memories experimentally demonstrated so far
have ``lifetimes'' of the order of 100 ns.
Thus, an event rate of $10^{7}$ Hz would effectively lead to an on-demand source.
Assuming a broadband source of especially Gaussian resource states,
considering our conditional GKP state generation scheme, a 10 GHz repetition rate or equivalently $10^{10}$ pulses per second would be feasible.
A success probability for the conditional state generation of the order of $10^{-3}$ would then result in a quasi-on-demand GKP qubit source.
Our calculations suggest that GKP state fidelities around 0.995 are possible for such values of the success probability, though assuming an ideal, loss-free scheme.
We also presented a short loss analysis, which implies demanding loss thresholds, but confirms the in-principle functioning of our scheme.
In order to fully realize such a scheme, the elementary single-mode cubic phase gates could be incorporated quasi-on-demand by combining cubic phase gate teleportation techniques
with all-optical quantum memories of $\sim 100$ ns lifetime.

Considering the advances of experimental single-mode cubic phase gates \cite{Fcpgexp}, it is hoped that these results may lead the way towards an experimental realization of optical GKP states as well as other applications.
As all four relevant Hamiltonians of our schemes are experimentally available in trapped-ion and superconducting circuit platforms, there is a multitude of existing applications, which the presented approximations could help bring into the optical context. The scope of this hybrid gate decomposition scheme thus reaches far beyond the creation of optical GKP states or two-qubit two-photon entangling gates.

\begin{acknowledgements}
    We acknowledge funding from the BMBF in Germany (QR.X, PhotonQ, QuKuK, QuaPhySI), from the EU/BMBF via QuantERA (ShoQC), from the EU's HORIZON Research and Innovation Actions (\mbox{CLUSTEC}), and from the Deutsche Forschungsgemeinschaft (DFG, German Research Foundation) TRR 306 QuCoLiMa (“Quantum Cooperativity of Light and Matter”).
\end{acknowledgements}

\onecolumngrid
\appendix

\section{Calculations}\label{AppCalculations}

\subsection{Representing the operators \texorpdfstring{$S_{\bm{\lambda}}(t)$}{S(t)} and \texorpdfstring{$T_{\bm{\lambda}}(t)$}{T(t)} by the polynomials \texorpdfstring{$P_{xx}$}{Pxx}, \texorpdfstring{$P_{xp}$}{Pxp}, \texorpdfstring{$P_{px}$}{Ppx}, and \texorpdfstring{$P_{pp}$}{Ppp}}

Given the unitary operator $\hat U$ and the set of quadratures $\mathbb{Q}=\{\hat x_1, \hat p_1, ..., \hat x_n, \hat p_n\}$ the operator $\hat U$ is well defined by its impact on the different quadratures $\hat U\hat q\hat U^{\dagger}$, $\hat q \in \mathbb{Q}$ up to a global phase. This can easily be seen by regarding the operator $\hat V$ with $\hat V\hat q\hat V^{\dagger}=\hat U\hat q\hat U^{\dagger}$ for all $\hat q \in \mathbb{Q}$. Then $\hat q=\hat U^\dagger\hat U\hat q\hat U^\dagger\hat U = \hat U^\dagger\hat V\hat q\hat V^\dagger\hat U$ and it follows that $[\hat q, \hat U^\dagger\hat V]=0$ for all $\hat q \in \mathbb{Q}$. Thus it must be $\hat U^\dagger\hat V=e^{i\phi}\thinspace \openone$.
Calculating the impact of the operator $S_{\bm{\lambda}}(t)$ on the four quadratures, we obtain
\begin{align}\begin{split}\label{Squads}
		S_{\bm{\lambda}}(t)\hat x_1S_{\bm{\lambda}}(-t)&=\hat x_1,\\
		S_{\bm{\lambda}}(t)\hat p_1S_{\bm{\lambda}}(-t)&=\hat p_1 - P_1[t\hat x_1]\hat x_2^2/2 - P_2[t\hat x_1]\hat p_2^2/2 - P_3[t\hat x_1](\hat x_2\hat p_2 + \hat p_2\hat x_2)/2,\\
		S_{\bm{\lambda}}(t)\hat x_2S_{\bm{\lambda}}(-t)&=P_{xx}[t\hat x_1]\hat x_2 + P_{xp}[t\hat x_1]\hat p_2,\\
		S_{\bm{\lambda}}(t)\hat p_2S_{\bm{\lambda}}(-t)&=P_{px}[t\hat x_1]\hat x_2 + P_{pp}[t\hat x_1]\hat p_2,
\end{split}\end{align}
with the polynomials $P_{xx}$, $P_{xp}$, $P_{px}$, $P_{pp}$, $P_1$, $P_2$, and $P_3$. 
The former can be calculated recursively given the relations
\begin{align}
	P_{xx}^{(0)}[t]&=1,\quad P_{xp}^{(0)}[t]=0,
	&P_{pp}^{(0)}[t]&=1,\quad P_{px}^{(0)}[t]=0,\nonumber\\
	P_{xx}^{(n)}[t]&=(1-\lambda_n\mu_nt^2)\thinspace P_{xx}^{(n-1)}[t] - \lambda_nt\thinspace P_{xp}^{(n-1)}[t],
	&P_{pp}^{(n)}[t]&=P_{pp}^{(n-1)}[t] + \mu_nt\thinspace P_{px}^{(n-1)}[t],\\
	P_{xp}^{(n)}[t]&=P_{xp}^{(n-1)}[t] + \mu_nt\thinspace P_{xx}^{(n-1)}[t],
	&P_{px}^{(n)}[t]&=(1-\lambda_n\mu_nt^2)\thinspace P_{px}^{(n-1)}[t] - \lambda_nt\thinspace P_{pp}^{(n-1)}[t].\nonumber
\end{align}
On the other hand, the latter are given by
\begin{align}
	P_1[t]&=(\partial_tP_{xx}[t])\thinspace P_{px}[t] - P_{xx}[t]\thinspace (\partial_tP_{px}[t]),\nonumber\\
	P_2[t]&=(\partial_tP_{xp}[t])\thinspace P_{pp}[t] - P_{xp}[t]\thinspace (\partial_tP_{pp}[t]),\\
	P_3[t]&=(\partial_tP_{xx}[t])\thinspace P_{pp}[t] - P_{xp}[t]\thinspace (\partial_tP_{px}[t]).\nonumber
\end{align}
This can be verified by comparing the recursion formulas of both sides of the equations while using the relation
\begin{align}\label{sincosinerelation2}
	P_{xx}[t]P_{pp}[t] - P_{xp}[t]P_{px}[t] &= 1\quad \forall t\in\mathbb{R}.
\end{align}
The impact of the operator $T_{\bm{\lambda}}(t)$ on the six quadratures is given by
\begin{align}
	T_{\bm{\lambda}}(t)\hat x_1T_{\bm{\lambda}}(-t)&=\hat x_1,&&
	T_{\bm{\lambda}}(t)\hat p_1T_{\bm{\lambda}}(-t)=\hat p_1 - P_1[t\hat x_1]\hat x_2\hat x_3 - P_2[t\hat x_1]\hat p_2\hat p_3 - P_3[t\hat x_1](\hat x_2\hat p_2 + \hat p_3\hat x_3),\nonumber\\
	T_{\bm{\lambda}}(t)\hat x_2T_{\bm{\lambda}}(-t)&=P_{xx}[t\hat x_1]\hat x_2 + P_{xp}[t\hat x_1]\hat p_3,&&
	T_{\bm{\lambda}}(t)\hat p_2T_{\bm{\lambda}}(-t)=P_{px}[t\hat x_1]\hat x_3 + P_{pp}[t\hat x_1]\hat p_2,\\
	T_{\bm{\lambda}}(t)\hat x_3T_{\bm{\lambda}}(-t)&=P_{xx}[t\hat x_1]\hat x_3 + P_{xp}[t\hat x_1]\hat p_2,&&
	T_{\bm{\lambda}}(t)\hat p_3T_{\bm{\lambda}}(-t)=P_{px}[t\hat x_1]\hat x_2 + P_{pp}[t\hat x_1]\hat p_3.\nonumber
\end{align}
Therefore the operators $S_{\bm{\lambda}}(t)$ as well as $T_{\bm{\lambda}}(t)$ are both well defined by the four polynomials $P_{xx}$, $P_{xp}$, $P_{px}$, and $P_{pp}$. Using this alternative representation will simplify the following calculations.

\subsection{Impact of \texorpdfstring{$S_{\bm{\lambda}}(t)$}{S(t)} on different input states}

Here we calculate the impact of $S_{\bm{\lambda}}(t)$ on the general two-mode input
\begin{align}
	\left|in\right> = \left(\hat S(-\xi)\left|\text{vac}\right>\right)_1\left(\hat S(-\zeta)\hat D\left(\alpha/\sqrt{2}\right)\left|\text{vac}\right>\right)_2,
\end{align}
with the squeezing operator $\hat S(\xi) = \exp\Big(\frac{1}{2}\left(\xi^*\hat a^2-\xi\hat a^{\dagger2}\right)\Big)$ and the displacement operator $\hat D(\alpha) = \exp(\alpha\hat a^\dagger - \alpha^*\hat a)$.
Therefore we consider the two differential equations
given by
\begin{align}
	0=&S_{\bm{\lambda}}(t)\left(\frac{\hat x_1}{k_1} + ik_1\hat p_1\right) S_{\bm{\lambda}}(-t)\thinspace S_{\bm{\lambda}}(t)\left|in\right>,\label{smode1}\\
	0=&S_{\bm{\lambda}}(t)\left(\frac{\hat x_2}{k_2} + ik_2\hat p_2 - \alpha\right) S_{\bm{\lambda}}(-t)\thinspace S_{\bm{\lambda}}(t)\left|in\right>\label{smode2}
\end{align}
when taking $\hat p_1\rightarrow -i\partial_{x_1}$ and $\hat p_2\rightarrow -i\partial_{x_2}$ in the position basis representation.
Here, $k_1$ and $k_2$ are given by $k_1(\xi)=\left(\frac{\cosh(|\xi|)+\sinh(|\xi|)\thinspace\xi/|\xi|}{\cosh(|\xi|)-\sinh(|\xi|)\thinspace\xi/|\xi|}\right)^{\frac{1}{2}}$ and $k_2(\zeta)=\left(\frac{\cosh(|\zeta|)+\sinh(|\zeta|)\thinspace\zeta/|\zeta|}{\cosh(|\zeta|)-\sinh(|\zeta|)\thinspace\zeta/|\zeta|}\right)^{\frac{1}{2}}$. Solving Eq.~\eqref{smode2} using the relations of Eq.~\eqref{Squads}, then employing Eq.~\eqref{smode1} and finally normalizing the result, we find that
\begin{align}
	\bra{x_1, x_2}S_{\bm{\lambda}}(t)\left|in\right> &= \frac{e^{-\frac{\alpha_I^2}{2}}}{\sqrt{k_1\pi}\sqrt{B}}\exp\left(-\frac{x_1^2}{2k_1^2}-\frac{A}{2B}x_2^2+\frac{\alpha}{B}x_2 - \frac{\alpha^2}{2}\frac{k_2P_{pp}[tx_1]}{B}\right),
\end{align}
with $A=P_{xx}[tx_1]/k_2 + ik_2P_{px}[tx_1]$ and $B=k_2P_{pp}[tx_1] - iP_{xp}[tx_1]/k_2$.
Here the notation $\sqrt{B}$ is used for the solution to the differential equation
\begin{align}
	\frac{f'(x)}{f(x)}=\frac{1}{2}\frac{B'(x)}{B(x)}.
\end{align}
The imaginary phase of the function $\sqrt{B}(x)$ thus covers the full range of $(-\pi, \pi]$ instead of the common $(-\frac{\pi}{2}, \frac{\pi}{2}]$.
Moreover, the results are only fixed up to a global phase.

When setting $\zeta=0$ and applying a Fourier transform in $x_2$ we arrive at Eq.~\eqref{GKPstate} of the main text. On the other hand, setting $\xi = \alpha = 0$ and using
\begin{align}
	&S_{\bm{\lambda}}(t)\left|\text{vac}\right>_1\left(\hat S(-\zeta)\left|1\right>\right)_2=S_{\bm{\lambda}}(t)\hat S_2(-\zeta)\hat a^\dagger_2\hat S_2(\zeta)S_{\bm{\lambda}}(-t)\thinspace S_{\bm{\lambda}}(t)\left|in\right>= S_{\bm{\lambda}}(t)\frac{\sqrt{2}\hat x_2}{k_2} S_{\bm{\lambda}}(-t)\thinspace S_{\bm{\lambda}}(t)\left|in\right>
	=\sqrt{2}S_{\bm{\lambda}}(t)\label{n0ton1}\\
	&\times\left(\frac{\hat x_2}{k_2}+\frac{iP_{xp}[tx_1]}{k_2B}\left(\frac{\hat x_2}{k_2}+ik_2\hat p_2\right)\right) S_{\bm{\lambda}}(-t)\thinspace S_{\bm{\lambda}}(t)\left|in\right>=\sqrt{2}\frac{\hat x_2}{k_2}\left(P_{xx}[tx_1] + iP_{xp}[tx_1]\frac{A}{B}\right) \thinspace S_{\bm{\lambda}}(t)\left|in\right>
	=\frac{\sqrt{2}\hat x_2}{B}\thinspace S_{\bm{\lambda}}(t)\left|in\right>\nonumber
\end{align}
gives us
\begin{align}
	\bra{x_a, y}S_{\bm{\lambda}}(t)\left|\text{vac}\right>_1\left(\hat S(-\zeta)\left|n_0\right>\right)_2 = \frac{\pi^{-\frac{1}{2}}}{\sqrt{ B}}\thinspace\left(\frac{\sqrt{2}y}{B}\right)^{n_0}\thinspace\exp\left(-\frac{A}{B}\frac{y^2}{2}-\frac{x_a^2}{2}\right).
\end{align}
with $n_0=0,1$.
Applying the operator $\hat M^{(j)}=e^{-i\hat x_a/2}S^{(a, j)}_{\bm{\lambda}}(\sqrt{\pi})$ and a Fourier transform in $x_a$ four times to two qumode states and an ancilla
\begin{align}
	\left|\psi_{n_y, n_z}\right>=\hat M^{(1)}\hat F^\dagger_a \hat M^{(2)}\hat F^\dagger_a \hat M^{(1)}\hat F^\dagger_a \hat M^{(2)}\hat F^\dagger_a\left|0\right>_a\left|n_y\right>\left|n_z\right>,
\end{align}
we get
\begin{align}\begin{split}
	\psi_{n_y, n_z}(x_a, y, z)&=
	e^{-ix_a/2}\int\frac{dp_a}{\sqrt{2\pi}}e^{ip_ax_a+ip_a/2} \int\frac{dx'_a}{\sqrt{2\pi}}e^{-ix'_ap_a+ix'_a/2} \int\frac{dp'_a}{\sqrt{2\pi}}e^{ip'_ax'_a-ip'_a/2}
	\left(\frac{\sqrt{2}y}{B}\right)^{n_y}\left(\frac{\sqrt{2}z}{D}\right)^{n_z}\\
	&\times
	\frac{\pi^{-\frac{3}{4}}}{\sqrt{B\thinspace D}}
	\exp\left(-\frac{p_a'^{2}}{2}-\frac{A}{B}\frac{y^2}{2}-\frac{C}{D}\frac{z^2}{2}\right),\label{CNOTFourier}
\end{split}\end{align}
with
\begin{align}\begin{split}
	A&=P_{xx}[\sqrt{\pi}x_a](P_{xx}[\sqrt{\pi}x'_a]-iP_{px}[\sqrt{\pi}x'_a])+iP_{px}[\sqrt{\pi}x_a](P_{pp}[\sqrt{\pi}x'_a]+iP_{xp}[\sqrt{\pi}x'_a]),\\
	B&=P_{pp}[\sqrt{\pi}x_a](P_{pp}[\sqrt{\pi}x'_a]+iP_{xp}[\sqrt{\pi}x'_a])-iP_{xp}[\sqrt{\pi}x_a](P_{xx}[\sqrt{\pi}x'_a]-iP_{px}[\sqrt{\pi}x'_a]),\\
	C&=P_{xx}[\sqrt{\pi}p_a](P_{xx}[\sqrt{\pi}p'_a]+iP_{px}[\sqrt{\pi}p'_a])-iP_{px}[\sqrt{\pi}p_a](P_{pp}[\sqrt{\pi}p'_a]-iP_{xp}[\sqrt{\pi}p'_a]),\\
	D&=P_{pp}[\sqrt{\pi}p_a](P_{pp}[\sqrt{\pi}p'_a]-iP_{xp}[\sqrt{\pi}p'_a])+iP_{xp}[\sqrt{\pi}p_a](P_{xx}[\sqrt{\pi}p'_a]+iP_{px}[\sqrt{\pi}p'_a]).
\end{split}\end{align}
The three Fourier transforms of Eq.~\eqref{CNOTFourier} can then be done numerically to obtain the output state of the approximate CZ gate.

\subsection{Impact of \texorpdfstring{$T_{\bm{\lambda}}(t)$}{T(t)} on different input states}

Next we are going to calculate the impact of $T_{\bm{\lambda}}(t)$ on the general three-mode input
\begin{align}
	\left|in\right> = N\thinspace\hspace{-1.mm}\int\hspace{-1.8mm}dx\hspace{-1.mm} \int\hspace{-1.8mm}dy\hspace{-1.mm} \int\hspace{-1.8mm}dz\ \phi(x)\exp\left(-\lambda\frac{y^2}{2} -\mu\frac{z^2}{2} - i\rho yz\right)\left|x\right>_1\left|y\right>_2\left|z\right>_3.
\end{align}
As $T_{\bm{\lambda}}(t)$ has no impact on $\phi(x)$, we can set $\phi(x)=\exp(-x^2/2)$ without loss of generality. This gives us three differential equations arising from the position space representation of
\begin{align}
	0=&T_{\bm{\lambda}}(t)\left(\hat x + i\hat p\right) T_{\bm{\lambda}}(-t)\thinspace T_{\bm{\lambda}}(t)\left|in\right>,\\
	0=&T_{\bm{\lambda}}(t)\left(\lambda \hat y + i\hat q + i\rho\hat z\right) T_{\bm{\lambda}}(-t)\thinspace T_{\bm{\lambda}}(t)\left|in\right>,\\
	0=&T_{\bm{\lambda}}(t)\left(\mu \hat z + i\hat r + i\rho\hat y\right) T_{\bm{\lambda}}(-t)\thinspace T_{\bm{\lambda}}(t)\left|in\right>,
\end{align}
with the pairs of quadratures $(\hat x, \hat p)$, $(\hat y, \hat q)$, and $(\hat z, \hat r)$. The solution is given by
\begin{align}
	\left<x, y, z|in\right> = \frac{N}{\sqrt{A}}\thinspace\phi(x)\exp\left(-\lambda\frac{y^2}{2A} -\mu\frac{z^2}{2A} - i\frac{B}{A}yz\right),
\end{align}
with $A=\left(P_{pp}[tx]+\rho P_{xp}[tx]\right)^2 + \lambda\mu P^2_{xp}[tx]$ and $B=\left(P_{pp}[tx]+\rho P_{xp}[tx]\right)\left(P_{px}[tx]+\rho P_{xx}[tx]\right) + \lambda\mu P_{xp}[tx]P_{xx}[tx]$. Similar to Eq.~\eqref{n0ton1} we also obtain
\begin{align}
	T_{\bm{\lambda}}(t)\left(\alpha\hat y + \beta\hat z\right)\left|in\right>= \left(\frac{\alpha\left(P_{pp}[tx]+\rho P_{xp}[tx]\right) + i\beta\lambda P_{xp}[tx]}{A}\hat y
	+ \frac{\beta\left(P_{pp}[tx]+\rho P_{xp}[tx]\right) + i\alpha\mu P_{xp}[tx]}{A}\hat z\right)T_{\bm{\lambda}}(t)\left|in\right>.
\end{align}
When setting $u_k=0$, one step of the protocol of Eq.~\eqref{HastrupProtocol} is given by
\begin{align}
	\left|\psi_{v_k, w_k}\right> = \hat F_3T_{\bm{\lambda}}(w_k)\hat F_3^\dagger\hat F_1 T_{\bm{\lambda}}(v_k)\hat F_1^\dagger \left(\alpha\hat y + \beta\hat z\right)\left|in\right>.
\end{align}
Besides the two operators $T_{\bm{\lambda}}(v_k)$ and $T_{\bm{\lambda}}(w_k)$ the Fourier transforms in $z$ can also be calculated analytically. Overall this leaves us with
\begin{align}
	\psi_{v_k, w_k}(x, y, z)=
	N'\thinspace\hspace{-1.mm}\int\hspace{-1.mm}\frac{dp}{\sqrt{2\pi}}e^{ipx} \int\hspace{-1.mm}\frac{dx'}{\sqrt{2\pi}}e^{-ix'p} 
	\left(\alpha'\hat y + \beta'\hat z\right)
	\phi(x')\exp\left(-\lambda'\frac{y^2}{2} -\mu'\frac{z^2}{2} - i\rho' yz\right),
\end{align}
where
\begin{align}
	\lambda'&={\textstyle \frac{\mu}{A[v_kp]}\left[ \left(P_{px}[w_kx]-i\frac{B[v_kp]}{\mu}P_{xx}[w_kx]\right)^2 + \left(\frac{\lambda}{\mu}+\frac{B^2[v_kp]}{\mu^2}\right)P_{xx}^2[w_kx]\right],}\nonumber\\
	\mu'&={\textstyle \frac{\mu}{A[v_kp]} \left[ \left(P_{pp}[w_kx]-i\frac{B[v_kp]}{\mu}P_{xp}[w_kx]\right)^2 + \left(\frac{\lambda}{\mu}+\frac{B^2[v_kp]}{\mu^2}\right)P_{xp}^2[w_kx]\right],}\\
	\rho'&={\textstyle i\frac{\mu}{A[v_kp]}\left[ \left(P_{pp}[w_kx]-i\frac{B[v_kp]}{\mu}P_{xp}[w_kx]\right)\left(P_{px}[w_kx]-i\frac{B[v_kp]}{\mu}P_{xx}[w_kx]\right) + \left(\frac{\lambda}{\mu}+\frac{B^2[v_kp]}{\mu^2}\right)P_{xp}[w_kx]P_{xx}[w_kx]\right],}\nonumber\\
	\alpha'&={\textstyle P_{xx}[w_kx]\frac{\alpha\left(P_{pp}[v_kp]+\rho P_{xp}[v_kp]\right) + i\beta\lambda P_{xp}[v_kp]}{A[v_kp]} - P_{px}[w_kx]\frac{\beta\left(P_{pp}[v_kp]+\rho P_{xp}[v_kp]\right) + i\alpha\mu P_{xp}[v_kp]}{A[v_kp]},}\nonumber\\
	\beta'&={\textstyle P_{pp}[w_kx]\frac{\beta\left(P_{pp}[v_kp]+\rho P_{xp}[v_kp]\right) + i\alpha\mu P_{xp}[v_kp]}{A[v_kp]} - P_{xp}[w_kx]\frac{\alpha\left(P_{pp}[v_kp]+\rho P_{xp}[v_kp]\right) + i\beta\lambda P_{xp}[v_kp]}{A[v_kp]}\quad\text{and}\quad N'=\frac{N}{\sqrt{A[v_kp]}}}.\nonumber
\end{align}
This step can be repeated for each iteration of the protocol. Calculating the Fourier transforms numerically then leaves us with the output state of the approximated measurement-free protocol.

\section{Optimized parameter sets}\label{AppOPS}

In order to improve upon the Trotter-Suzuki decomposition the parameter sets $\bm{\lambda}$ are separately optimized for each application. For this the Basin-hopping algorithm implemented in Python with starting points fulfilling Eq.~\eqref{conditionsOOC} is used. For the approximated CZ gate the worst-case fidelity defined in the main text is maximized. For the GKP states the state's fidelity towards the corresponding Gaussian GKP state is used as a figure of merit. The optimized gate sets approximating the CZ gate are

{\small
\begin{align*}
	&L = 6:& &[0.1917, 0.3068, 0.3478, 0.3615, 0.3199, 0.1972]\\
	&L = 7:& &[0.1453, 0.2971, 0.3045, 0.3139, 0.3057, 0.2976, 0.1485, 0]\\
	&L = 9:& &[0.1163, 0.2294, 0.2462, 0.2440, 0.2304, 0.2442, 0.2464, 0.2302, 0.1168, 0]\\
	&L = 11:& &[0.09535, 0.1918, 0.1981, 0.1946, 0.1926, 0.1985, 0.1926, 0.1946, 0.1980, 0.1920, 0.09504, 0]\\
	&L = 13:& &[0.08059, 0.1627, 0.1651, 0.1630, 0.1636, 0.1649, 0.1623, 0.1649, 0.1635, 0.1630, 0.1650, 0.1628, 0.08046, 0]\\
	&L = 15:& &[0.06972, 0.1404, 0.1418, 0.1404, 0.1413, 0.1414, 0.1403, 0.1420, 0.1402, 0.1414, 0.1412, 0.1404, 0.1418, 0.1405, 0.06964, 0]
\end{align*}
}The optimized gate sets used for the probabilistic GKP creation scheme are
{\small
	\begin{align*}
		&&&\mathclap{\text{Square encoding}\ \ \ \ \ \ \ \ \ \ \ \ \;}\\
		&L = 3:& &[0.6794, 0.4543, 0.3353, 0]\\
		&L = 5:& &[0.5217, 0.3469, 0.2937, 0.2536, 0.1937, 0]\\
		&L = 7:& &[0.3566, 0.2243, 0.2416, 0.2731, 0.2764, 0.2306, 0.1295, 0]\\
		&L = 9:& &[0.02422, 0.7957, 0.4211, 0.2941, 0.2488, 0.2294, 0.2202, 0.1823, 0.08733, 0]\\
		&&&\mathclap{\text{Qunaught encoding}\ \ \ \ \ \ \ \ \ \:}\\
		&L = 9:& &[0.3433, 0.1839, 0.1593, 0.1827, 0.2088, 0.2074, 0.2037, 0.1716, 0.09903, 0]\\
		&&&\mathclap{\text{Hexagonal encoding}\ \ \ \ \ \ \ \ \ }\\
		&L = 9:& &[-0.05847, -0.2111,  0.5222,  0.3127,  0.2532, 0.2260, 0.2096, 0.1735, 0.08916, 0]\\
		&&&\mathclap{\text{Magic state}\ \ \ \ \ \ \ \ \ \ \ \ \ \ \ \ \ \ \ \ }\\
		&L = 16:& &[0.1038, 0.07294, 0.1861, 0.1610, 0.08781, 0.09327, 0.1317, 0.1501, 0.1380,
		0.1348, 0.1213, 0.1173, 0.1105,\\
		&&&\ 0.1096, 0.09671, 0.05178]\\
		&L = 21:& &[0.04155, 0.08769, 0.1025, 0.05514, 0.1391, 0.1291, 0.06734, 0.07781 , 0.1304,
		0.1295, 0.09831, 0.07139, 0.08922,\\
		&&&\ 0.1150, 0.1115, 0.08428, 0.07268, 0.09360,
		0.1119, 0.09423, 0.04103, 0]\\
		&L = 25:& &[0.02913, 0.09474, 0.1146, 0.05894, 0.09529, 0.09499, 0.06886, 0.07248, 0.09725,
		0.09047, 0.07591, 0.07003, 0.08284,\\
		&&&\ 0.08626, 0.08380, 0.07685, 0.07693, 0.07845,
		0.08287, 0.08220, 0.07977, 0.07979, 0.08801, 0.07813, 0.03718, 0]
	\end{align*}
}For the deterministic GKP creation scheme the optimized gate set is
{\small
	\begin{align*}
		&L = 11:& &[0.09506, 0.1881, 0.1951, 0.1963 , 0.1907, 0.1945, 0.1919 , 0.1943, 0.1972, 0.1870, 0.09451, 0]\qquad\qquad\qquad\qquad\qquad\quad
	\end{align*}
}

\noindent The gates of the schematic circuit of Fig.~\ref{Schema} are given by
\begin{align}
	Sq = \hat S\left(-\ln(k)\right),\quad D_1=\hat D\left(\frac{kd}{\sqrt{2}\pi}\right),\quad D_2=\hat D\left(-i\frac{k^2d}{2\sqrt{2}\pi}\right),\quad X^3=\exp\left(ir_j\hat x^3\right),
\end{align}
with $d=2\sqrt{\pi}$ and $k=\sqrt{\frac{21}{4}\pi}$. The $r_j$'s are dependent on the beam splitters and can thus be adapted. Following Eq.~\eqref{decomp1} and choosing $s=\arccos\left(\frac{1}{\sqrt{3}}\right)$ leads to the weakest possible cubic phase gates with $r_1=r_2=0.0643$, $r_3=r_4=0.0872$, $r_5=r_6=0.1304$, and $r_7=-0.1085$. On the other hand, setting $s_1\neq s_2\neq s_3$ allows us to obtain $r_1=r_2=r_3=r_4=r_5=r_6=-r_7=0.1675$.

\twocolumngrid
\bibliography{AllOpticalQCUsingCPG}

\end{document}